\PassOptionsToPackage{pdfpagelabels=false}{hyperref}
\documentclass[useAMS,fleqn,usenatbib]{mnras}
\pdfoutput=1 
\usepackage{graphicx}
\usepackage{amsmath,amssymb,amstext}
\usepackage[T1]{fontenc}
\usepackage{ae,aecompl}
\usepackage[utf8]{inputenc}
\usepackage{newtxtext,newtxmath}
\usepackage[figure,figure*,table,table*]{hypcap}
\usepackage[normalem]{ulem}
\usepackage[dvipsnames]{xcolor}
\usepackage{subcaption}
\usepackage{enumitem}
\usepackage{appendix}
\include{hyperlink-year-only-natbib-patch}
\pdfmapfile{+rsfso.map}
\DeclareSymbolFont{rsfso}{U}{rsfso}{m}{n}
\DeclareSymbolFontAlphabet{\mathscr}{rsfso}

\let\origitem\item
\renewcommand{\item}{\normalfont\origitem}
\newcommand{\bolditem}{\normalfont\bfseries\origitem}

\newcommand*{\sub}[2]{\ensuremath{{#1}_{#2}}}

\newcommand*{\kms}{\text{km}\,\text{s}\ensuremath{^{-1}}}
\newcommand*{\msun}{\sub{\text{M}}{\odot}}
\newcommand*{\perh}{\ensuremath{h^{-1}}}

\newcommand*{\chisq}{\ensuremath{\chi^2}}

\newcommand*{\mailto}[1]{\href{mailto:#1}{#1}}
\newcommand*{\http}[1]{\href{http://#1}{#1}}
\newcommand*{\https}[1]{\href{https://#1}{#1}}

\title[Illuminating Dark Matter Halo Density Profiles Without Subhaloes]{Illuminating Dark Matter Halo Density Profiles Without Subhaloes}

\author[C.~E.~Fielder et al.]{%
Catherine~E.~Fielder,$^{1,2}$\thanks{E-mail: \mailto{cef41@pitt.edu}}
Yao-Yuan~Mao,$^{1,2,3}$\thanks{E-mail: \mailto{yymao.astro@gmail.com}; NHFP Einstein Fellow}
Andrew~R.~Zentner,$^{1,2}$ 
Jeffrey~A.~Newman,$^{1,2}$\newauthor
Hao-Yi~Wu,$^{4}$ and
Risa~H.~Wechsler$^{5,6}$
\vspace*{6pt}
\\
$^{1}$Department of Physics and Astronomy, University of Pittsburgh, Pittsburgh, PA 15260, USA\\
$^{2}$Pittsburgh Particle Physics, Astrophysics, and Cosmology Center (PITT PACC), University of Pittsburgh, Pittsburgh, PA 15260, USA\\
$^{3}$Department of Physics and Astronomy, Rutgers, The State University of New Jersey, Piscataway, NJ 08854, USA\\
$^{4}$Center for Cosmology and Astro-Particle Physics, The Ohio State University, Columbus, OH 43210, USA\\
$^{5}$Kavli Institute for Particle Astrophysics and Cosmology; Physics Department, Stanford University, Stanford, CA 94305, USA\\
$^{6}$SLAC National Accelerator Laboratory, Menlo Park, CA 94025, USA
\vspace*{-12pt}
}

\pubyear{2020}
\begin{document}

\label{firstpage}
\pagerange{\pageref{firstpage}--\pageref{lastpage}}
\maketitle

\begin{abstract}
Cold dark matter haloes consist of a relatively smooth dark matter component as well as a system of bound subhaloes. It is the prevailing practice to include all mass, including mass in subhaloes, in studies of halo density profiles in simulations. However, often in observational studies satellites are treated as having their own distinct dark matter density profiles in addition to the profile of the host. This difference can make comparisons between theoretical and observed results difficult. In this work we investigate density profiles of the smooth components of host haloes by excluding mass contained within subhaloes. We find that the density profiles of the smooth halo component (without subhaloes) differ substantially from the conventional halo density profile, declining more rapidly at large radii. We also find that concentrations derived from smooth density profiles exhibit less scatter at fixed mass and a weaker mass dependence than standard concentrations. Both smooth and standard halo profiles can be described by a generalised Einasto profile, an Einasto profile with a modified central slope, with smaller residuals than either an NFW or Einasto profile. These results hold for both Milky Way-mass and cluster-mass haloes. This new characterisation of smooth halo profiles can be useful for many analyses, such as lensing and dark matter annihilation, in which the smooth and clumpy components of a halo should be accounted for separately.

\end{abstract}


\section{Introduction}
\label{Section:into}

The gravitational collapse of over-dense patches 
of the universe in the Lambda--Cold Dark Matter ($\Lambda$CDM) cosmology culminates in the formation of virialized dark matter-dominated structures called dark matter haloes. Much of the modelling necessary to undertake modern data analyses for the purposes of understanding cosmology, the nature of dark matter, or the evolution of galaxies relies on our understanding of various properties of dark matter haloes. Halo properties and their evolution have been investigated theoretically and in great detail
\citep[e.g.,][]{white1991,kauffmann1993,cole1994,somerville1999,benson2000,zheng2005,guo2013,ludlow2019}. Similarly there are a number of observational probes of the structures of dark matter haloes, including gravitational lensing \citep{meneghetti2005,mandelbaum2006a,comerford2007,johnston2007,okabe2013,umetsu2016}, X-ray surface brightness/temperature \citep{nagai2007}, and line-of-sight velocity dispersion of either stars \citep{battaglia2005,newman2013,akin2016} or satellites \citep{merritt1987,more2011}. The delicate marriage of dark matter halo theory and observation is fundamental for our understanding of galaxy formation and evolution.

$N$-body simulations have been instrumental in determining the structures and mass assembly histories of haloes. Haloes grow hierarchically. Large haloes accumulate their masses through mergers with smaller haloes, which leave a lasting imprint. On dynamical timescales ($\sim$3--4\,Gyr at $z=0$), infalling haloes may be stripped of their mass and/or disrupted, while those that survive become dense, self-bound objects orbiting within the larger host halo 
\citep[see e.g.,][]{kauffmann1993,zentner2003,zentner2005,bullock2010}. 
Therefore, host dark matter haloes are composed of 
(1) a relatively smooth component and 
(2) \emph{subhaloes}, the gravitationally bound 
haloes that reside within the larger hosts. The smooth component of a halo consists primarily of disrupted subhaloes from earlier mergers \citep[e.g.,][]{purcell2007,zavala2019}, but there may also be a sub-dominant contribution acquired via smooth accretion \citep[e.g.,][]{wang2011}. 

A basic characteristic of a dark matter halo is its density profile, which describes the way that mass is distributed throughout the bound object. Despite the fact that halo formation is a complex process, $N$-body simulations have shown that the mass distributions of dark matter haloes are well described by the same equilibrium density profile at all masses. The most widely used halo density profile is the isotropic Navarro--Frenk--White (NFW) profile \citep{nfw1996,nfw1997}. The NFW profile is a two-parameter (one-parameter at fixed mass) functional form that describes the halo mass distribution as a function of distance from the halo centre:
\begin{equation}
    \rho(r) = \frac{\rho_{s}}{\frac{r}{r_s}(1+\frac{r}{r_s})^{2}},
\label{eq:nfw}
\end{equation} 
where $\rho_{s}$ is the characteristic over-density. The length scale, $r_{s}$, is referred to as the characteristic radius or scale radius; it corresponds to the radius at which $\mathrm{d}\ln \rho/\mathrm{d}\ln r = -2$.

The degree to which the mass within a halo is concentrated toward the halo centre is often quantified by the dimensionless halo concentration parameter, $c = \frac{r_{\rm vir}}{r_{s}}$. The virial radius, $r_{\rm vir}$, can be thought of as the  ``outer edge'' of the halo that  characterises the halo size. Halo concentration has a well-known dependence upon both mass (the concentration--mass relation) and redshift \citep[e.g.,][]{nfw1997,bullock2001,klypin2016}. Profile parameters of individual haloes, such as concentration, are typically determined by building a spherically averaged density profile from simulation data and fitting this profile to the NFW form. 

The contributions of substructure are handled differently in observational and theoretical studies, which provides challenges to compare results between them. In simulation analyses, spherical averages of haloes typically include all halo substructure. In observational investigations, such as the analyses of gravitational lens systems, the spherically averaged halo profile is typically modelled with a particular form, the parameters of which are then inferred from the data \citep[e.g.,][]{moller2002,limousin2006}. However, in such observational studies, it is often the case that the observed substructures (e.g., satellite galaxies) within a lensed system are assigned their own, distinct density profiles in addition to the profile attributed to the host system \citep[e.g.,][]{newman2013,nierenberg2017,despali2018,gilman2020}. Thus there arises the risk that these subhaloes are being double counted. If the subhaloes are to be treated as discrete units with their own properties, one needs to use halo density profiles developed to exclude the mass within substructures. Likewise, if one were to exclude subhaloes from the host halo and then model the subhaloes separately, there is no guarantee that the same \emph{host} halo mass profile (e.g., NFW) would still be an accurate representation of the host. These concerns are the focus of the present work.

In this work we examine the impact of subhaloes on host halo mass profiles, and the ability of analytic profiles like the NFW profile to accurately describe halo mass distributions that do not include the mass associated with subhaloes.
To do this, we make use of the density profiles of haloes from which we have removed the mass associated with subhaloes, a technique first used by \citet{wu2013}. Even when we exclude subhaloes we are still spherically averaging the haloes. In order to explore a wide range of masses, we use two sets of high-resolution zoom-in simulations. One simulation focuses on Milky Way-mass haloes and the other on cluster-sized haloes.

In \autoref{Section:data} we describe the simulations used in our work, the halo finder, and the halo density calculations. \autoref{Section:Analysis} details our analysis methods of the halo profiles through statistical methods and fitting to different mass profile prescriptions. \autoref{section:results} presents the results for both stacked and individual haloes in order to compare different functional forms for the density profile. In \autoref{Section:Conclusion} we summarise our results and conclude that the mass distributions of the smooth component of the halo and the combination of the smooth and subhalo components are distinctly different, especially in the outer region of the halo mass distribution and describe the implications of such a result with a more universal mass profile, the generalised Einasto profile.

\section{Simulations}
\label{Section:data}

In the following section, we outline the simulations and halo finder used,  the methods for calculating simulated halo density profiles, and the procedures used to excluding mass associated with subhaloes.

\subsection{Zoom-In Simulations of Two Mass Regimes}
\label{Subsection:simulations}

In this work we use two sets of zoom-in, gravity-only simulations. The first set consists of 45 zoom-in simulations focusing on haloes of approximately the mass of the Milky Way's halo and initially presented in \citet{mao2015}, which we refer to as the Mao et al. Milky way Mass Zoom-in (MMMZ) simulations in the remainder of this paper. The second simulation suite comprises 96 cluster-mass zoom-in simulations known as the RHAPSODY simulations, first presented in \citet{wu2013}. All analyses performed in this paper are applied to both sets of simulations, allowing us to test for consistency of results across several orders of magnitude in halo mass and to test for mass dependence. In both cases we use the present-day ($z=0$) snapshots.

The haloes in the MMMZ simulations cover a very narrow range in virial mass ($M_{\rm vir} =  10^{12.1 \pm 0.03}\ \mathrm{M}_{\odot}$). The cosmological parameters for the simulations are as follows: matter density $\Omega_{\rm M} = 0.286$; dark energy density $\Omega_{\Lambda} = 1 - \Omega_{\rm M} = 0.714$; Hubble parameter $h = H_0 / 100 = 0.7$; mass fluctuation amplitude $\sigma_{8} = 0.82$; and scalar spectral index $n_{\rm s} = 0.96$.

These MMMZ haloes were selected for high-resolution re-simulation from a parent \texttt{c125-1024} dark-matter-only simulation run with L-GADGET \citep[see][]{becker2015,lehmann2017}.
The high-resolution zoom-in regions have a particle mass of $m_{\rm p} = 3.0 \times 10^{5}\;\perh\mathrm{M}_{\odot}$ and a softening length of 170 \perh\,pc comoving. We take the resolution limit in our analyses to be four times the softening length, or $0.68$ \perh\,kpc for the Milky Way mass haloes. The lower limit to the subhalo maximum circular velocity $V_{\rm max}$ for convergence is approximately 10\,\kms. For more details on the MMMZ simulations, refer to \citet{mao2015}. 

The RHAPSODY simulations also span a very narrow mass range at the cluster-size scale, $M_{\rm vir} = 10^{14.8\pm0.05}\ \mathrm{M}_{\odot}$. The cosmological parameters used for the RHAPSODY simulations are very similar to those of MMMZ, specifically: $\Omega_{\rm M} = 0.25$; 
$\Omega_{\Lambda} = 0.75$; 
$h=0.7$; $\sigma_{8} = 0.8$; and $n_{\rm s} = 1$. 
The differences between MMMZ and RHAPSODY cosmologies 
will have negligible effect on any comparisons performed in this paper, as halo concentrations only depend weakly on cosmological parameters, especially at $z=0$ \citep[e.g.,][]{ludlow2014}. 

These RHAPSODY zoom-in simulations were selected from one of the CARMEN simulations from the LArge Suite of DArk MAtter Sim-ulations (LasDamas; \citealt{mcbride2011}) with a volume of $1$ \perh Gpc and $1120^{3}$ particles. We use the higher-resolution version of these simulations 
(RHAPSODY 8K), which have a particle mass of $m_{\rm p} = 1.3 \times 10^{8} \perh\mathrm{M}_{\odot}$ and a force resolution (as defined above as four times the softening length) equivalent to $13$ \perh\,kpc.

\subsection{Density Profiles and Subhalo Removal}
\label{subsection:calc}

In this work, we endeavour to study the density distribution of the ``smooth" components of host haloes. We use \textsc{Rockstar} (version 0.99.9-RC3+) to identify haloes according to a virial definition. The \textsc{Rockstar} halo finder uses phase-space information in order to distinguish subhaloes from the host halo's background density, which we take advantage of for this work. Using the particle catalogues we define the {\em smooth} component of any host halo 
to be the mass not associated with any \textsc{Rockstar}-identified subhalo. For more details on \textsc{Rockstar},
see \citet{behroozi2013} or access the publicly available code at \https{bitbucket.org/gfcstanford/rockstar}.

There exists no unambiguous way to define the smooth component of a halo. Our operational definition is dependent upon the \textsc{Rockstar} algorithm for halo identification. Further details regarding this choice and other possible ambiguities are given in \autoref{subsection:consistency} and Appendix~\ref{Section:rockstar_sub_removal}.

For the purposes of this work, we define the smooth 
component of a host halo to be those particles not 
explicitly associated with a self-bound subhalo by 
\textsc{Rockstar}. We have used this definition to 
construct the following processed simulation data sets.
\begin{itemize}
    \item \textit{subhalo-included:} the set of all particles associated with the host and all of its subhaloes (in \textsc{Rockstar} this corresponds 
    to all particles that reside within the virial 
    radius of the host halo). \\
    \item \textit{subhalo-excluded:} particles that are associated with the host but \textit{not} explicitly 
    associated with any subhalo listed in the \textsc{Rockstar} halo catalogue. This sample 
    excludes subhaloes but includes diffuse substructures, 
    such as streams, that do not meet the 
    binding criteria in \textsc{Rockstar}. \\
    \item \textit{subhalo-only:} the set of particles 
    explicitly associated with at least one subhalo. This corresponds to the set of particles in the subhalo-included set that are not within the subhalo-excluded set.\\
\end{itemize}
In all cases, only particles within the virial radius of the host halo are considered. As is evident, the 
\textit{subhalo-included} data set consists of the 
union of particles within the \textit{subhalo-excluded} 
and \textit{subhalo-only} data sets, and the \textit{subhalo-excluded} 
and \textit{subhalo-only} data sets are mutually exclusive.
Readers interested in further details of these 
definitions can see 
Appendix~\ref{Section:rockstar_sub_removal}.

\begin{figure*}
    \centering
    \includegraphics[width=\textwidth]{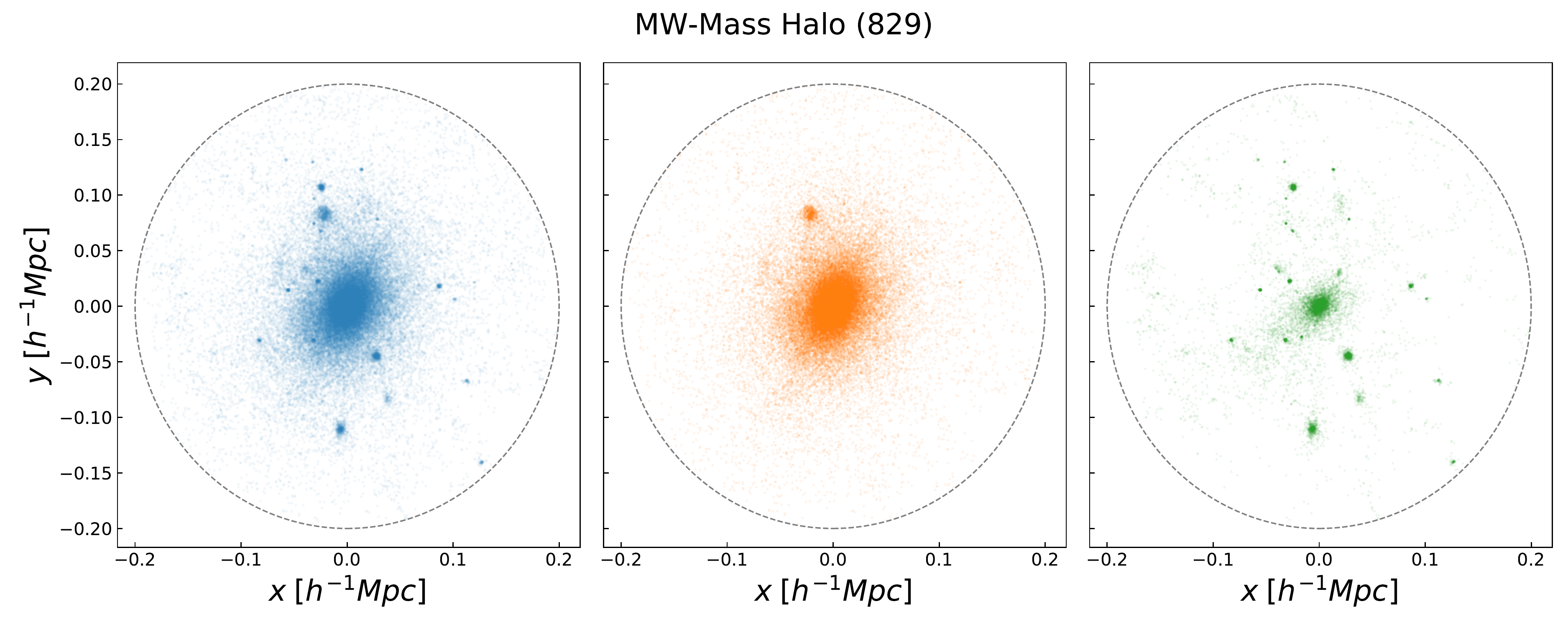}
    \caption{Visualisation of subhalo removal in a 2D projection for one of the MMMZ host haloes. The particles shown here are all restricted to be within the virial radius of the halo, denoted by the black dashed line. In addition we plot a subset of the particles restricted to be near the $z$ value of the halo centre, requiring $z_{\rm{host}}-0.001 < z < z_{\rm{host}}+0.001 \perh \rm{Mpc}$. The left (blue) panel shows the standard case where all host and subhalo particles are considered together as making up the dark matter halo. High density areas away from the centre corresponding to subhaloes can be easily seen. The middle (orange) panel corresponds to the case where we have excluded subhalo particles; his case may be more appropriate for comparison with observational techniques, as discussed in \autoref{Section:data}. The particle distribution is much smoother and more uniform, but some substructure (not associated with systems that \textsc{Rockstar} defines as self-bound subhaloes) is still visible. In the right panel (green) we show only those particles associated with subhaloes that are included in the \textsc{Rockstar} catalogue. By construction the orange distribution and the green distribution added together correspond to the blue slice.}
    \label{fig:vizualization}
\end{figure*}

It is useful to get a visual impression of the substructure removal procedure. 
\autoref{fig:vizualization} shows a scatter plot of particles within a plane of width $\Delta z = \pm 1$kpc of the 
centre of the host halo for one of the MMMZ haloes. The host halo is centred at $(x=0,y=0)$. The left (blue) panel shows the combined distribution of host halo and subhalo particles; i.e., the {\em subhalo-included} sample as defined above. This is the set of particles that would generally be taken to correspond to a single dark matter halo in $N$-body simulation analyses and in the vast majority of studies of halo properties. 

The middle (orange) panel of \autoref{fig:vizualization} 
depicts the {\em subhalo-excluded} sample. 
There is a noticeable feature aside from the 
smooth central component of the host halo; 
at $(x=-0.01, y=0.09 \perh \rm{Mpc})$ 
there is an over-density that was not 
listed in the \textsc{Rockstar} halo catalogue due to its low bound fraction (the subhalo is not self-bound in the eyes of \textsc{Rockstar}). This definition of halo particles more 
closely maps to the dark matter associated with 
the central galaxy of the halo, and hence is closer to the 
\emph{de facto} halo definition assumed in many 
observational analyses.

For comparison, the right (green) panel depicts only the particles associated with those subhaloes that are listed in the \textsc{Rockstar} halo table, or {\em subhalo-only}. By construction, the combination of the orange and green points is identical to the blue points. While it appears that there is a remaining central halo component, these particles exhibit a large, coherent velocity to the halo centre when observed in velocity space. This over-density represents a substructure passing near the host halo centre. 

It is evident that the distribution of mass without subhaloes is much smoother with many fewer density peaks, demonstrating the effectiveness of our subhalo exclusion procedure.

The smooth, {\em subhalo-excluded}, 
components of halo profiles contain less total mass 
than the virial masses of the full halo profiles. 
For instance, for the MMMZ haloes, $\sim16\%$ of the mass 
is in subhaloes on average, 
while the mass fraction in subhaloes 
for the RHAPSODY haloes is $\sim 42\%$. 
Consequently, the smooth components of the haloes will 
not separately satisfy the same virial over-density 
criterion that the {\em subhalo-included} profiles satisfy. 
This introduces an ambiguity in 
the physical boundaries of the haloes with 
subhaloes removed. We choose to use 
the original virial radius, $r_{\rm vir}$, as the 
halo boundary even in the case with subhaloes 
removed.  We note that quantities 
listed with the subscript "nosub" are 
computed when subhaloes 
are \emph{excluded}. We also note that when individual haloes are examined we include their snapshot reference ID.

\section{Analysis}
\label{Section:Analysis} 

In this section, we describe our analysis methods. 
First, we summarise the functional forms of the 
analytic dark matter halo profiles used to fit the simulation data, 
including a new functional form that we refer to as the 
``generalised Einasto'' profile. 
We then describe the algorithms used 
to fit these profiles to simulated halo density profiles.  Lastly, we specify the statistics used to assess 
the quality of fits. 

\subsection{Density Profile Parameterisation}
\label{Subsection:methods}

Halo density profiles have been modelled and 
parameterised in various different ways.
The two following families of functions have been used most 
frequently to describe dark matter halo densities:
\begin{enumerate}
    \item \textit{Double Power-Law:} These profiles asymptote to different power laws at small and large radii. Profiles of the double power-law type generally have functional forms similar to the NFW profile described by \autoref{eq:nfw}. Previous work has explored a generalised NFW (gNFW) five-parameter profile in order to account for deviations of dark matter haloes from the standard NFW profile, particularly to allow for variations in the slopes at the inner (cusp/core) and outer regions of haloes \citep{hernquist1990,zhao1996,dekel2017}. The gNFW profile can be written as
    \begin{equation}
        \rho(r) = 
        \frac{2^{(\beta-\gamma)/\alpha}\rho_{s}}
        {\left(\frac{r}{r_{s}}\right)^{\gamma}\left[1+\left(\frac{r}{r_{s}}\right)^{\alpha}\right]^{(\beta-\gamma)/\alpha}},
    \label{eq:gennfw}
    \end{equation}
    where $\alpha$, $\beta$, and $\gamma$ are all constants. The generalised NFW double power law has a slope of $-\gamma$ at small radii and $-\beta$ at large radii; the $\alpha$ parameter governs the rate of transition between these slopes. The standard NFW profile has  $(\alpha,\beta,\gamma) = (1,3,1)$, corresponding to an inner power law index of $-1$ and an outer power law index of $-3$. Many studies use the NFW profile, as the simplicity of having fewer free parameters can be 
    advantageous, despite the increased fidelity with which a gNFW profile can represent simulation data.\\
    
    \item \textit{Continuously Varying Power-Law:} Functional forms of this type allow for a gradual flattening of the slope of the density profile towards the inner part of the halo. A well-known member of the continuously varying power-law family is the Einasto profile \citep{einasto1963}, which has come to be used extensively 
    as recent work has shown it to be a better description of the distribution of matter in haloes than other two or three-parameter profiles \citep{gao2008,navarro2010,dutton2014,klypin2016}. The standard three-parameter Einasto profile is
    \begin{equation}
        \rho(r) = \rho_{s}\mathrm{exp}\Big(-\frac{2}{\alpha}\Big[\Big(\frac{r}{r_{s}}\Big)^{\alpha}-1\Big]\Big),
    \label{eq:einasto}
    \end{equation}
    where $\rho_{s}$ and $r_{s}$ are the scale density and scale radius, defined similarly as in the NFW case. The Einasto profile differs from the NFW profile most at small and large radii. The inner slope of the Einasto profile goes to 0 as $r$ approaches 0, and the outer slope does not asymptote to a constant value.
\end{enumerate}

In the main body of this paper, we focus on four specific profiles. These are (i) the NFW profile (\autoref{eq:nfw}, two parameters), (ii) the generalized NFW profile (\autoref{eq:gennfw}, five parameters), and (iii) the Einasto profile (\autoref{eq:einasto}, three parameters), which serve as standards to compare to prior work, as well as (iv) a new four-parameter ``generalised Einasto'' (gEinasto) profile that we find best describes halo density profiles:
\begin{equation}
\rho(r) = \rho_{s}\Big(\frac{r}{r_{s}}\Big)^{-\gamma}\mathrm{exp}\Big(-\frac{2}{\alpha}\Big[\Big(\frac{r}{r_{s}}\Big)^{\alpha}-1\Big]\Big),
\label{eq:geinasto}
\end{equation}
where $\gamma$ modifies the inner density profile slope compared to the \citep{einasto1963} form. This profile's asymptotic behaviour tends to $-\gamma$ at small radii\footnote{As we were finalising this manuscript, \citet{lazar2020} presented a cored-Einasto profile for the description of density profiles in galaxy formation simulations.}.

In Appendix~\ref{Section:other_profiles}
we describe the other functional forms that we 
tested, but do not present detailed results for each 
of these profiles for the sake 
of brevity. Our qualitative results 
carry over to these cases as well. 
As we will show, the 
gEinasto profile, {\bf (iv)}, is the best-performing 
model and will be used to demonstrate most of our 
results. We use the NFW 
model, {\bf (i)}, as a standard for comparison.

In order to enable comparisons across different types of halo profiles, we calculate halo concentration using the radius at which the profile fit has a derivative equal to $-2$, $r=r_{-2}$. We define concentration as 
$c_{-2} = r_\text{vir}/r_{-2}$.
For each of the aforementioned profiles, 
the local power-law indices and values of 
$r_{-2}$ are as follows.
\begin{enumerate}
    \bolditem The NFW profile:
        \begin{equation}
        \frac{\mathrm{d}\ln \rho(r)}{\mathrm{d}\ln r} = -\frac{3r+r_{s}}{r+r_{s}};
        \quad\quad
        r_{-2} = r_{s}.
        \label{eq:nfw_deriv}
\end{equation}
    \bolditem The generalised NFW profile: 
        \begin{equation}
            \frac{\mathrm{d}\ln \rho(r)}{\mathrm{d}\ln r} = -\gamma + \frac{(\gamma-\beta)(\frac{r}{r_{s}})^{\alpha}}{1+(\frac{r}{r_{s}})^{\alpha}};
            \quad\quad
            r_{-2} = r_{s}\frac{(\gamma-2)}{(2-\beta)}^{1/\alpha}.
            \label{eq:gnfw_deriv}
        \end{equation}
    \bolditem the Einasto profile
        \begin{equation}
        \frac{\mathrm{d}\ln \rho(r)}{\mathrm{d}\ln r} = -2\left(\frac{r}{r_{s}}\right)^{\alpha};
        \quad\quad
        r_{-2} = r_{s}.
        \label{eq:einasto_deriv}
        \end{equation}
    \bolditem The generalised Einasto profile
        \begin{equation}
        \frac{\mathrm{d}\ln \rho(r)}{\mathrm{d}\ln r} = -\gamma-2\left(\frac{r}{r_{s}}\right)^{\alpha};
        \quad\quad
        r_{-2} = r_{s}\left(\frac{2-\gamma}{2}\right)^{1/\alpha}.
        \label{eq:geinasto_deriv}
        \end{equation}
\end{enumerate}
 In the case of the gEinasto profile, if 
 $\gamma > 2$, then the profile will never 
 have $\mathrm{d}\ln \rho/\mathrm{d}\ln r \geq -2$.

\subsection{Profile Fitting Procedure}
\label{subsection:fitting}

Much of the analysis in this paper relies on fits of halo density profiles to the functional forms described in \autoref{Subsection:methods}. We perform fits on both stacked and individual halo profiles. Stacked profiles have compelling uses in both theory (individual halo profiles are noisy, making stacks useful for the study of general trends) and observation (e.g. the stacking of weak lenses to improve the inferrence of halo properties from low signal-to-noise lenses). Individual halo profiles allow us to study halo-to-halo variations. Our fitting procedure is as follows:
\begin{enumerate}
    \item From \textsc{Rockstar} catalogues and particle tables, we compute halo density profiles as follows. We bin the distribution of particle distances relative to the centre of the host halo, $r$, into 90 logarithmically spaced bins between $r/r_{\rm vir} = 10^{-3}$ and 1. We calculate the density of each bin by dividing mass by bin volume, $v_i = \frac{4}{3}\pi (r_{i+1}^3-r_i^3)$, where $i$ is the bin index.\\
    \item To compute stacked profiles, 
    we first normalise each profile by the total mass 
    of its \textit{subhalo-included} profile. 
    Normalisation is not necessary for individual halo fits.\\
    \item To construct a stack, we calculate the mean density of the number-weighted profiles, creating a number-weighted stack. Stacks are constructed in scaled units ($r/r_{\rm vir}$). We calculate 
    standard error of the density in each bin, 
    $\mathrm{SE}_{i} = \sigma_{i}/\sqrt{N}$, where $\sigma_{i}$ is the standard deviation of the density of the given radial bin across all haloes of the stack, and $N$ is the number of haloes in the stack. We use the standard error as the uncertainty in the fits.\\
    \item For individual halo profiles we do not need to calculate the mean. In these cases, we use $\sigma$, the standard deviation of the density of all the haloes in the respective simulation, as the 
    uncertainty.\\ 
    \item We then mask all bins with bin centres below four times the softening length of the simulation from being used in the fits ($4\times l_{\rm soft} = 0.68$ \perh\,kpc comoving for MMMZ and $4\times l_{\rm soft} = 13$ \perh\,kpc for RHAPSODY).\\
    \item Finally, we fit each functional form to the density profiles (whether derived from a stack or an individual halo) by minimizing the usual $\chi^2$ function [see Eq.~(\ref{eq:chi2}) below] using $\tt{scipy.optimize.curve\_fit}$ from the {\tt scipy} Python package \citep{scipy}. To expedite the fits, we include by-eye initial guesses for each free parameter. The fitting errors are those described in points (iii) and (iv). 
    We apply the following set of bounds on each parameter to avoid 
    unphysical solutions. 
        \begin{enumerate}
            \item $\rho_{s} > 0$,
            \item $0 < r_{s} < R_{\rm vir}$,
            \item $ 0 < \alpha < 5$,
            \item $0 < \beta < 10$,
            \item $0 < \gamma < 5$ for gNFW or
            $-5 < \gamma < 5$ for gEinasto.
        \end{enumerate}
\end{enumerate}
The details of the fitting procedure are as follows. First, we choose as the independent variable for each binned density $r_i = \sqrt{r_{\rm bin, outer}\times r_{\rm bin, inner}}$, the geometric mean of the radial bin edges. We evaluate the analytic density profiles at each of these values of $r_i$. We allow $\rho_{s}$ to be a free parameter in the fits, but one may choose instead to keep it fixed, which guarantees that the fitted profile will satisfy the integral definition of the halo virial mass. There are studies that have allowed $\rho_{s}$ to be a free parameters \citep[e.g.,][]{bullock2001,wechsler2002,dicintio2014} and studies that have not \citep[e.g.,][]{navarro2004,denhen2005,ludlow2016,child2018}.
Also note that when we plot stacked profiles we normalise the profile to the mean mass of the haloes (with subhaloes \textit{included}) so that all profiles are plotted in absolute units. For example, we plot $r^2 \rho(r)$ in units of M$_{\odot}$\,kpc$^{-1}$.\\

\begin{figure}
    \centering
    \includegraphics[width=0.98\linewidth]{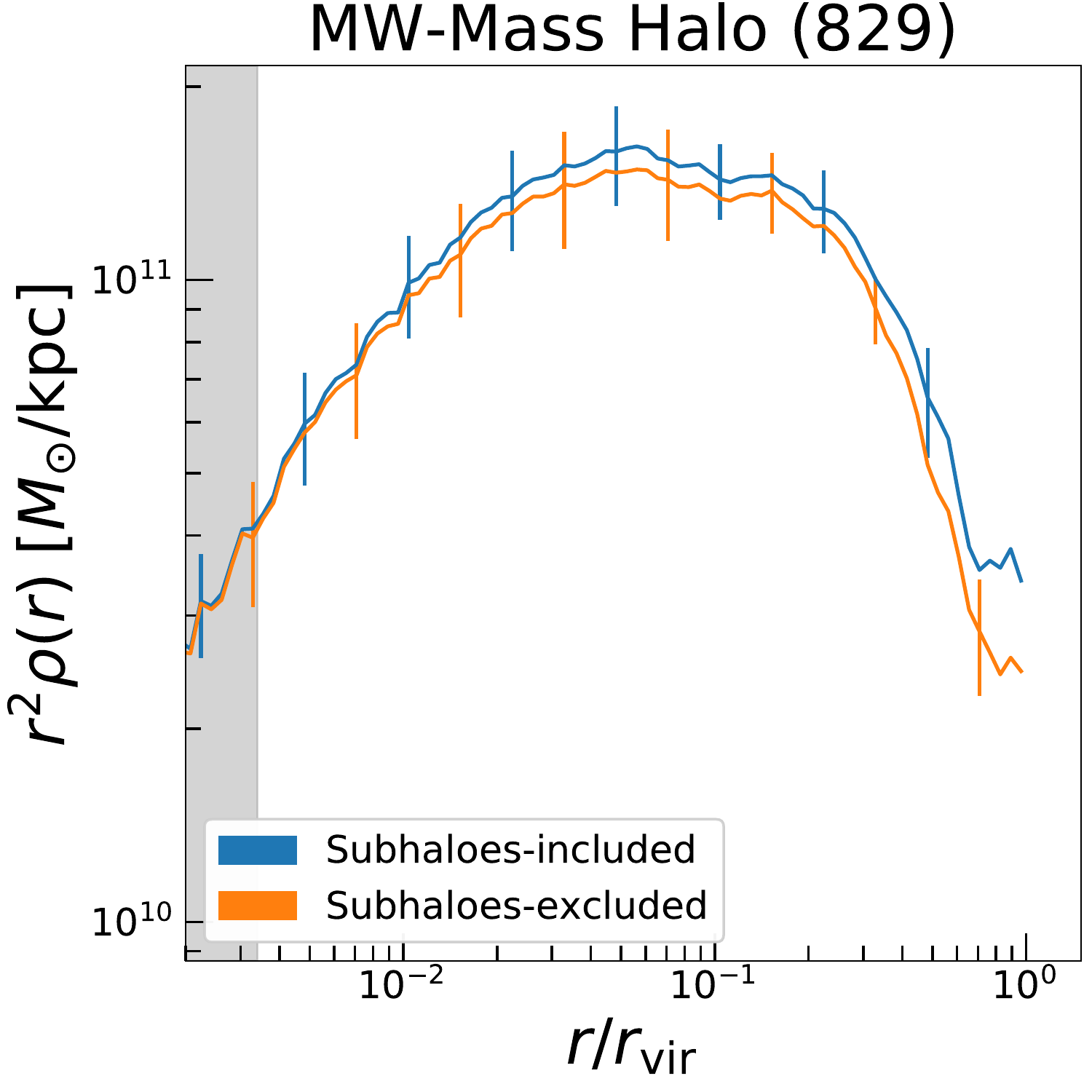}
    \caption{The density profile of Halo 829 in the Milky Way-mass (MMMZ) simulations, the same halo shown in \autoref{fig:vizualization}. The y-axis shows $r^{2}$ times the density, as a function of distance from the centre of the halo. The grey region represents radii below the resolution limit of the simulations, as discussed in \autoref{Section:data}. The blue and orange curves are based on particles shown in the correspondingly coloured panels in the previous figure; in all figures, blue indicates subhalo-included distributions and orange indicates subhalo-excluded distributions. The error bars on the simulation data are plotted every 10 bins starting at the first bin for subhalo-included and at the fifth bin for subhalo-excluded. These errors correspond to the halo-to-halo spread in density across all of the MMMZ haloes, rather than the uncertainty in the plotted values. In the inner region of the halo the two profiles are very similar to each other, but they deviate more in the outer parts of the halo; this is a common trend among haloes.}
    \label{fig:829profile}
\end{figure}

\autoref{fig:829profile} is the first depiction of a single halo's density profile in this paper and is shown as an example. The plot corresponds to MMMZ halo 829, the same object whose projected density is shown in \autoref{fig:vizualization}. On the y-axis we plot the quantity $r^{2} \times \rho(r)$ instead of just $\rho(r)$ to make it easier to see differences amongst density profiles, given the large dynamic range of density. The grey, vertical band at the left represents the resolution limit of four times the softening length, which is equivalent to $0.68$ \perh\,kpc comoving. The meanings of the colours are matched to \autoref{fig:vizualization}. The solid blue curve represents the subhalo-included density of Halo 829; i.e., the quantity normally calculated from $N$-body simulations, incorporating both host halo and subhalo mass. The solid orange curve depicts the subhalo-excluded density profile; mass associated with detected subhaloes is not counted in this case. While in the inner region of the halo the two profiles are quite similar they deviate from each other in the outer region. 
The error bars indicate the 
halo-to-halo scatter in density 
and not uncertainty in the density. 

As part of our analyses, we also scrutinise the local power-law index, 
${\rm d}\ln{\rho}/{\rm d}\ln{r}$, from both the numerical halo density profile data and the fits to the analytic profiles. Examining the local power-law index of the profiles yields insight into the profile parameters (e.g., $\alpha$, $\beta$, $\gamma$) that will best represent the numerical data. The derivatives of the numerical data are calculated numerically using the three-point (quadratic) Lagrangian interpolation estimator \citep{hildebrand1987,bevington2002}.

\begin{figure}
    \centering
    \includegraphics[width=0.98\linewidth]{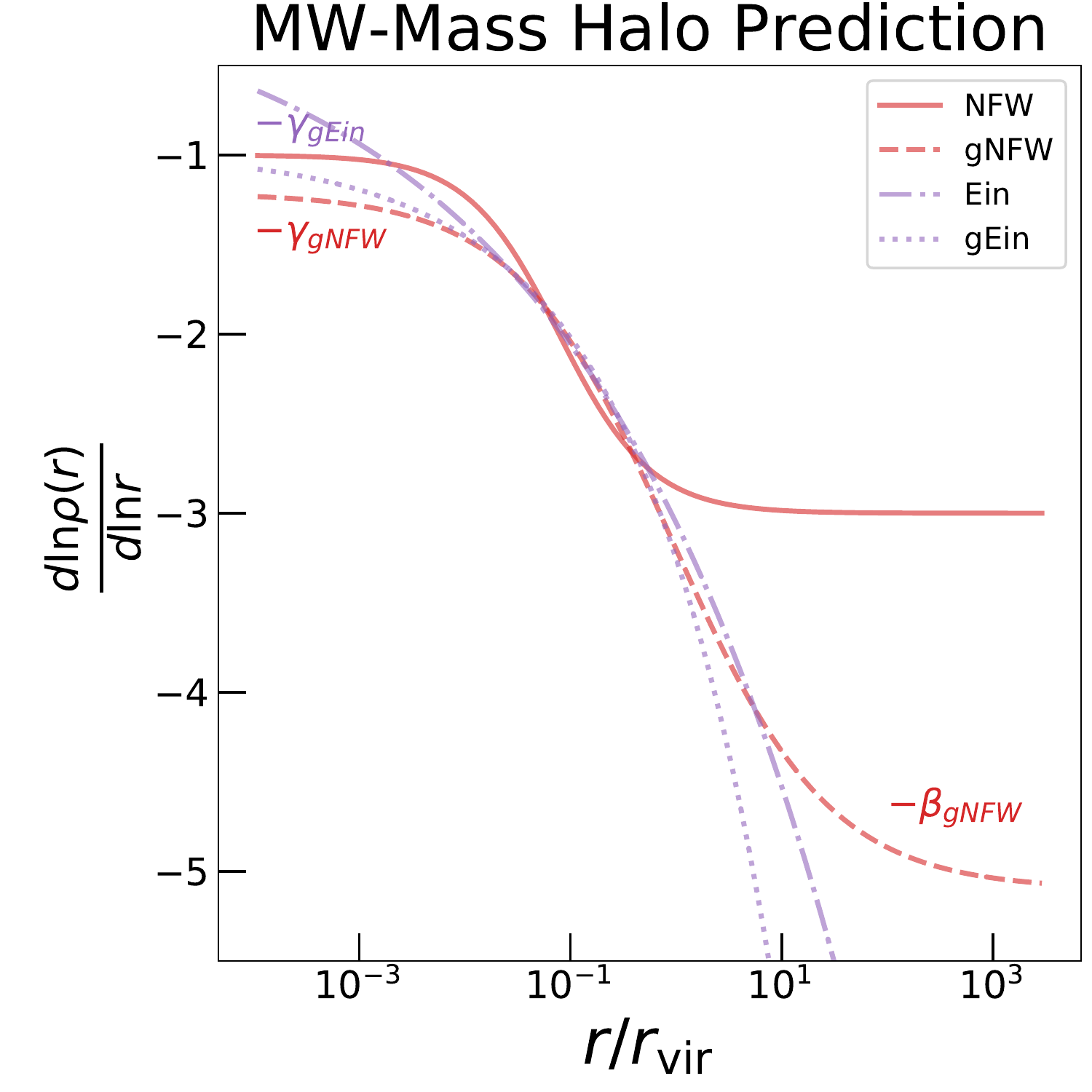}
    \caption{Analytic density profile derivatives for a Milky Way-mass (MMMZ) type halo. The parameters of each profile were determined from the subhalo-\textit{included} fits to the MMMZ stacked haloes  \autoref{Subsection:methods} (NFW, generalised NFW, Einasto, and generalised Einasto). This plot extends over a large range of radii in order to clearly depict clearly the asymptotic 
    behaviours of each profile. The red lines depict the NFW (solid) and generalised NFW (dashed) profiles, and the purple lines depict the Einasto (dash-dot) and generalised Einasto (dotted) profiles. Our parameterisation of the generalised Einasto profile allows for a shallower inner slope in the halo.}
    \label{fig:gen_deriv}
\end{figure}

In \autoref{fig:gen_deriv} we show an example of the local power law indices as a function of halo-centric distance for each of the analytic profiles that we investigate for pedagogical purposes. Red lines depict the NFW (solid) and generalised NFW (dashed) profiles, and purple lines depict the Einasto (dash-dotted) and generalised Einasto (dotted) profiles.  The parameters of these profiles were selected from fits to the {\em subhaloes-included} 
stacked profiles of the Milky Way-mass, MMMZ haloes and are listed in \autoref{tab:best_fit}, but the fits themselves are not important here. Notice that for illustrative purposes, the plot extends over an unusually large range of halo-centric distances. We do this in order to illustrate the asymptotic behaviours of the profiles, as each profile has unique behaviour at its asymptotes. The parameterization of the generalised Einasto profile allows us to decrease the asymptotic slope in the inner region of the halo.

\subsection{Assessing Fits}
\label{subsection:statistics}

We have employed a variety of statistics to assess the quality of fits for each functional form described in \autoref{Subsection:methods}. The statistics that we will examine are the root-mean-square fractional residual, \chisq, and Akaike and Bayesian information criteria. 

The first statistic we use to characterise fit accuracy is the root-mean-square fractional residual between a halo profile and its best fit, 
\begin{equation}
    \mathrm{fRMS} = \sqrt{\frac{1}{N_\mathrm{bins}} \sum_i \left[  
    \frac{ \rho_\mathrm{pred}(r_i) }{ \rho_\mathrm{data}(r_i) } - 1 \right]^2},
\label{eq:rms}
\end{equation}
where $\rho_{\rm data}(r_i)$ is the density estimated for 
the simulated halo in the $i^{\rm th}$ radial bin, 
$\rho_{\rm pred}(r_i)$ is the value of the density predicted by 
the model to which the data is being fit. 
The fRMS residual value provides an easily interpretable 
measure of the accuracy of fits; if $\mathrm{fRMS}=0.1$, 
the best fit of a given type typically 
deviates from the measured 
halo density profile by $\approx 1\%$ (in an RMS sense, 
so that deviations are added in quadrature). 
The closer fRMS is to zero, the better the match is between the 
best fit to a given functional form and the measured density profiles.

We use the well-known $\chi^2$ statistic to assess quality of fit and to perform model selection from among the analytic density 
profiles that we propose above. The $\chi^2$ statistic is 
\begin{equation}
    \chi^{2} = \sum_i 
    \left[ \frac{\rho_\mathrm{pred}(r_i) - \rho_\mathrm{data}(r_i) }{\mathrm{SE}_i} \right]^2,
    \label{eq:chi2}
\end{equation}
where $\mathrm{SE}_{i}$ is the standard error per bin described in \autoref{subsection:fitting}, and the sum over $i$ is a 
sum over all of the bins not excluded by the 
resolution cut. To avoid confusion, we denote $\chi^2$ values derived from fits to stacked density profiles as $\chisq_{\rm stack}$. 
For individual profiles, we calculate the $\chisq$ values 
resulting from each of the individual halo profile fits. 
In order to discuss the fit quality for this ensemble of 
fits, we examine the median $\chisq$ value resulting 
from the set of fits and designate 
it as $\chisq_{\rm median}$.

We aim to identify the profiles that best 
represent the halo profiles in our simulations; however, we cannot draw this conclusion directly from the fRMS and/or $\chi^2$ values because the functional 
forms used have varying numbers of free parameters. 
Adding additional parameters to a fitting function will 
always decrease both the minimum value of $\chi^2$ 
and fRMS. To assess whether or not the additional parameters 
have intrinsic explanatory power, it is necessary 
to determine whether or not the decrease in the minimum value of $\chi^2$ is significant in comparison to the 
decrease in $\chi^2$ one would expect from adding 
parameters that simply fit the noise in the data. 
The Akaike Information Criterion (AIC) and 
the Bayesian Information Criterion (BIC) 
are statistics that are frequently used to 
quantify whether or not the improvement of the fit 
(i.e., decrease in $\chi^2$) 
is sufficient to conclude that 
the additional parameters have explanatory power \citep{AIC2015,BIC2018}. 
The AIC and BIC are 
\begin{align}
\text{AIC} &\equiv 2{k} + \chi^{2}, \text{ and} \\
\text{BIC} &\equiv \ln({n}){k} + \chi^{2},
\end{align}
where $k$ is the number of free parameters 
in a given model and $n$ is the number of data points 
used to evaluate the fit. 
Differences in AIC/BIC of more than ten are 
generally considered to provide very strong evidence 
of a superior model. For example, if AIC is reduced by 
more than ten upon introducing a model with more parametric 
freedom, one concludes that the new model is a superior fit 
to the data with intrinsic explanatory power.

\section{Results}
\label{section:results}

In this section, we compare the subhaloes-included and -excluded halo density profiles, and assess how well each of the various analytic forms of halo density profile can describe the simulated data sets. In so doing, we will show that the subhaloes-included profiles are very different from the subhaloes-excluded density profiles, particularly in the outer regions of the halo, and characterise those differences. 

\subsection{Identifying Best Fit Forms for Stacked Profiles}
\label{subsection:stacks}

The focus of this subsection is to identify which analytic halo profile fitting functions best describe the stacked profiles of simulated dark matter haloes.

\autoref{tab:stats} presents a variety of summary statistics for assessing the fit quality for each of the proposed analytic density profiles introduced in \autoref{Subsection:methods}. 
Blue or orange colours are used to indicate tables of statistics for the subhalo-included or subhalo-excluded cases, respectively. All goodness-of-fit statistics provided are the values for fits to stacked halo profiles, with the exception of the last column, which lists the median chi-squared value from the set of fits to each individual halo in the sample. The profiles are listed in \autoref{tab:stats} in order of 
BIC from lowest (best) to highest (worst). It is evident that fRMS and $\chi^2$ generally 
follow the same rank ordering as BIC. As a reference, the best-fit values of the parameters from all of the profile fits are provided in \autoref{tab:best_fit} of Appendix~\ref{Section:best_fit_vals}.

The generalised Einasto profile is the most effective model for the simulated halo density profiles, considering both Milky-Way-mass haloes (MMMZ) and cluster-mass haloes (RHAPSODY) and for both the subhalo-included and subhalo-excluded data.
It is favoured over the other profiles considered by all statistics used here, except for the median chi-squared value across haloes.  In this case, with the exception of the the subhaloes-excluded MMMZ data, the generalised NFW profile yields marginally smaller values. The smaller \chisq values for the individual haloes are likely a result of the increased model complexity of the generalised NFW profile.

Having established the utility of the generalised Einasto profile, we will focus on this profile for most of the remainder of this paper. In most of the figures that follow, we will present detailed results for the gEinasto profile and the standard NFW profile. We include the standard NFW profile because it is widely studied and therefore enables comparison with previous literature. The corresponding rows in \autoref{tab:stats} are highlighted to indicate these profiles. 

\begin{table*}
\includegraphics[trim=5mm 5mm 5mm 15mm, clip, width=0.75\linewidth]{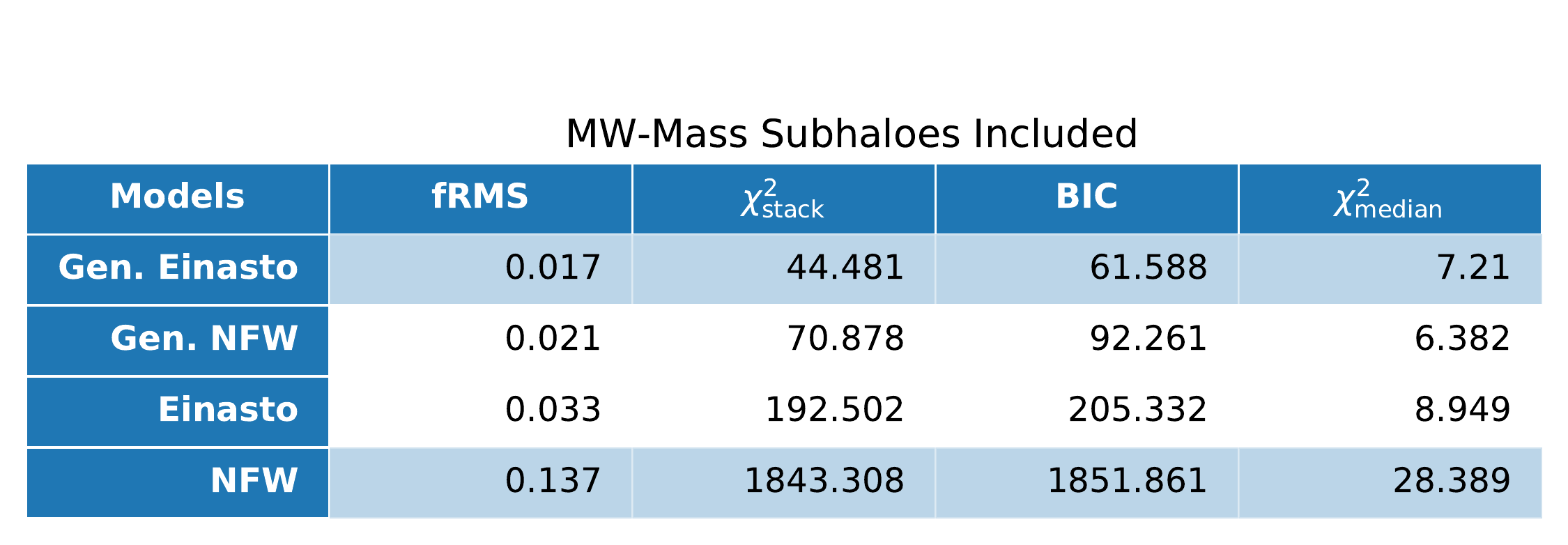}
\includegraphics[trim=5mm 5mm 5mm 15mm, clip, width=0.75\linewidth]{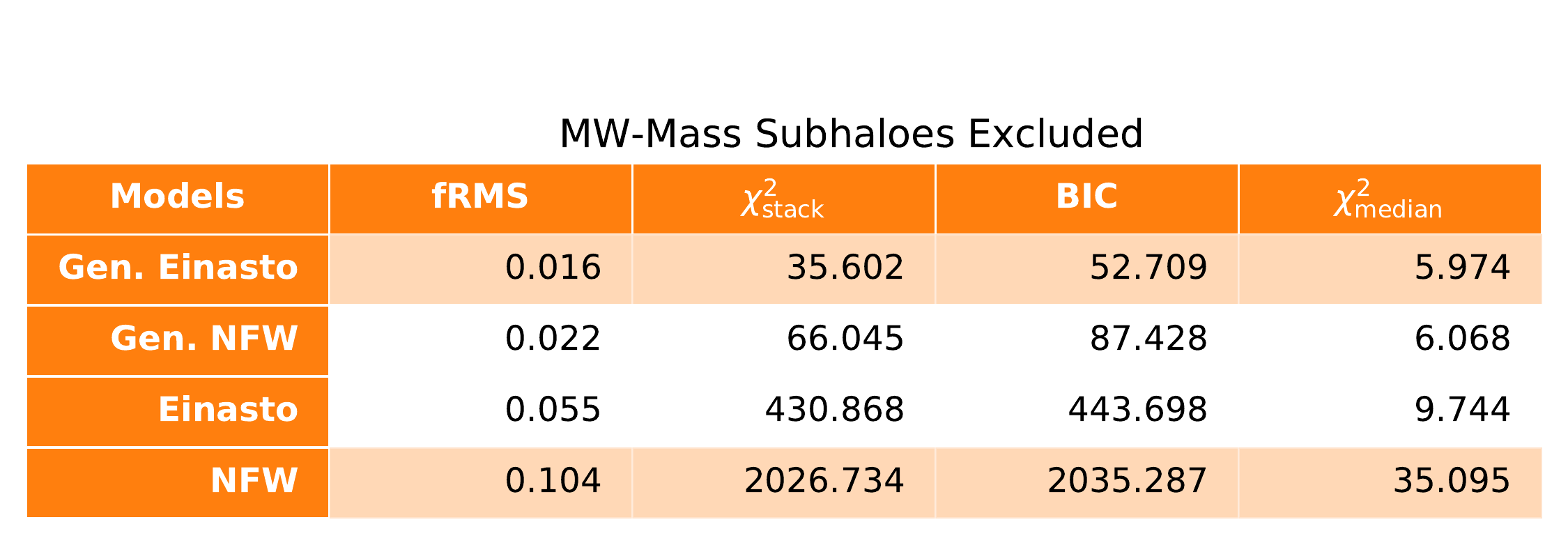}
\includegraphics[trim=5mm 5mm 5mm 15mm, clip, width=0.75\linewidth]{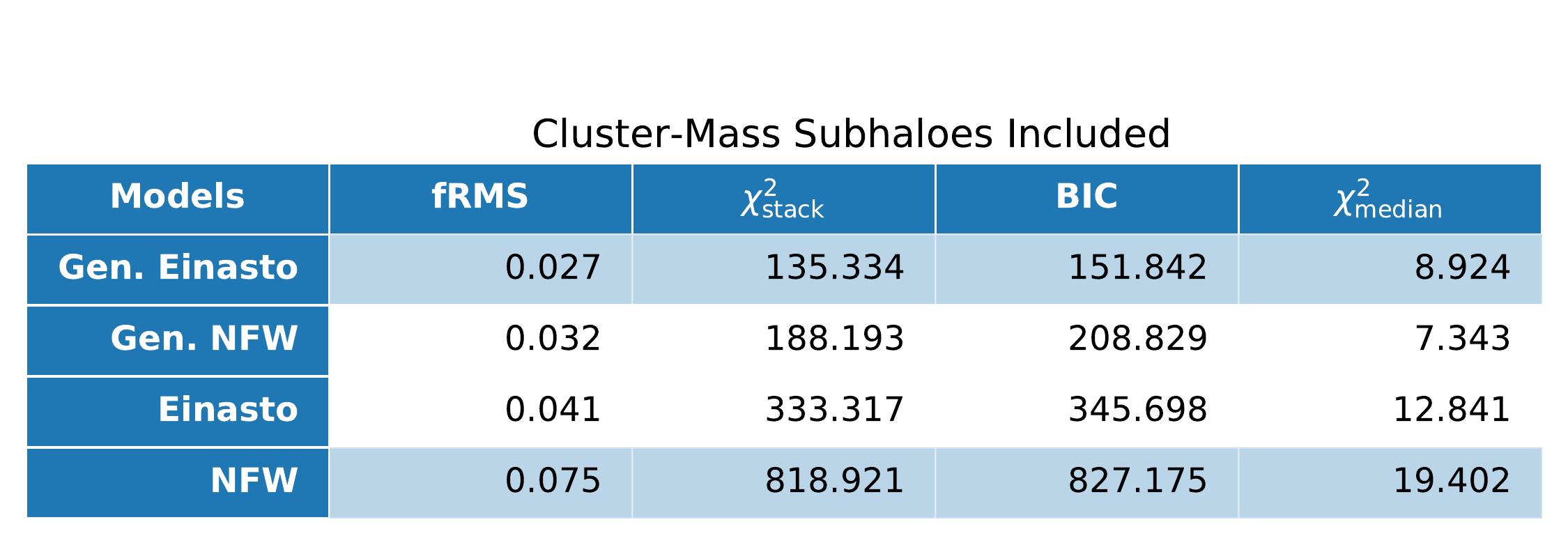}
\includegraphics[trim=5mm 5mm 5mm 15mm, clip, width=0.75\linewidth]{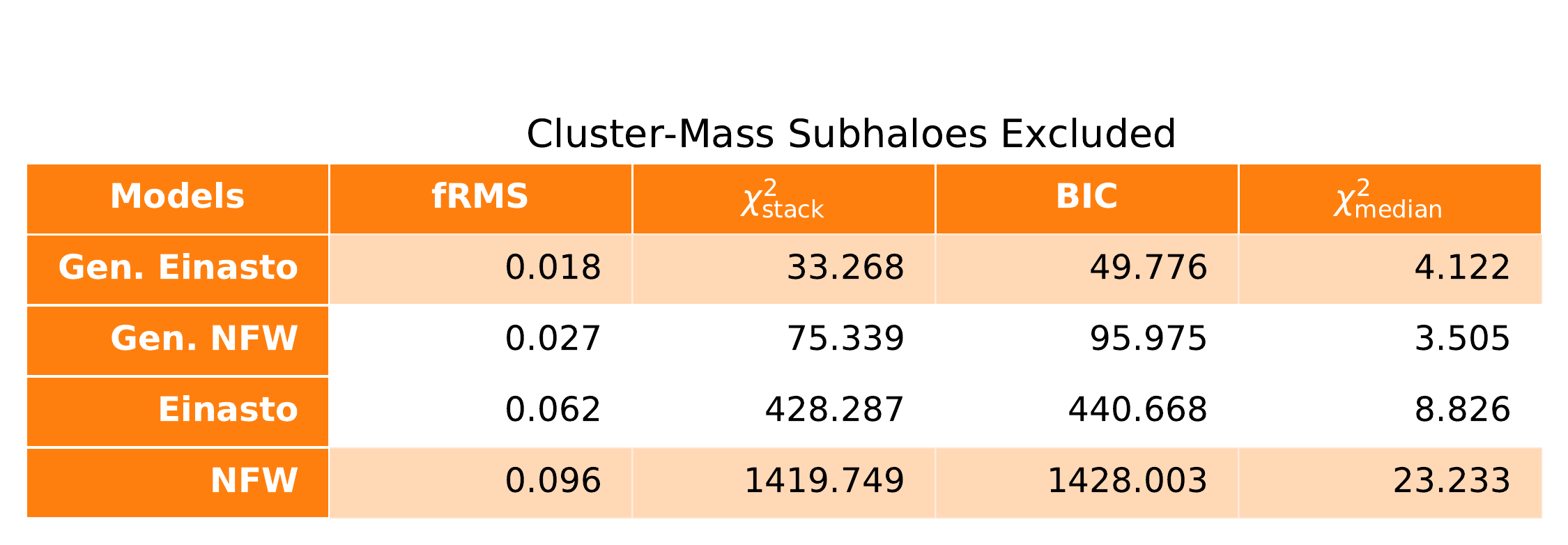}
\caption{Statistics for evaluating goodness-of-fit and model suitability for the NFW profile, generalised NFW profile, Einasto profile, and the generalised Einasto profile. The blue tables represent fits where mass from subhaloes is included in the density profile, as has generally been done in previous simulation studies; orange tables represent cases where subhalo mass has been excluded. The first two tables correspond to the Milky Way-mass (MMMZ) host haloes and the second two tables correspond to the cluster-mass (RHAPSODY) host haloes. The two  models used in the following figures are highlighted. We first list statistics calculated from fits to stacks (averages) of the haloes in each respective simulation; specifically, the the root mean square fractional residual (fRMS), the \chisq{} of the best fit to the stacked haloes ($\chisq_{\rm stack}$), and the Bayesian Information Criteria. The final column, $(\chisq_{\rm median})$, represents the median of the chi-squared values calculated from fits to the density profiles of all individual haloes from a simulation set. The statistics used are described in more detail in \autoref{subsection:statistics}. In almost all cases, the generalised Einasto profile provides the most balance of fit quality and limited model complexity; the BIC values in particular provide strong evidence that this functional form performs optimally at describing halo density profiles.}
\label{tab:stats}
\end{table*}
\medskip

\begin{figure*}
    \centering
    \begin{subfigure}{0.5\textwidth}
        \includegraphics[width=0.9\textwidth]{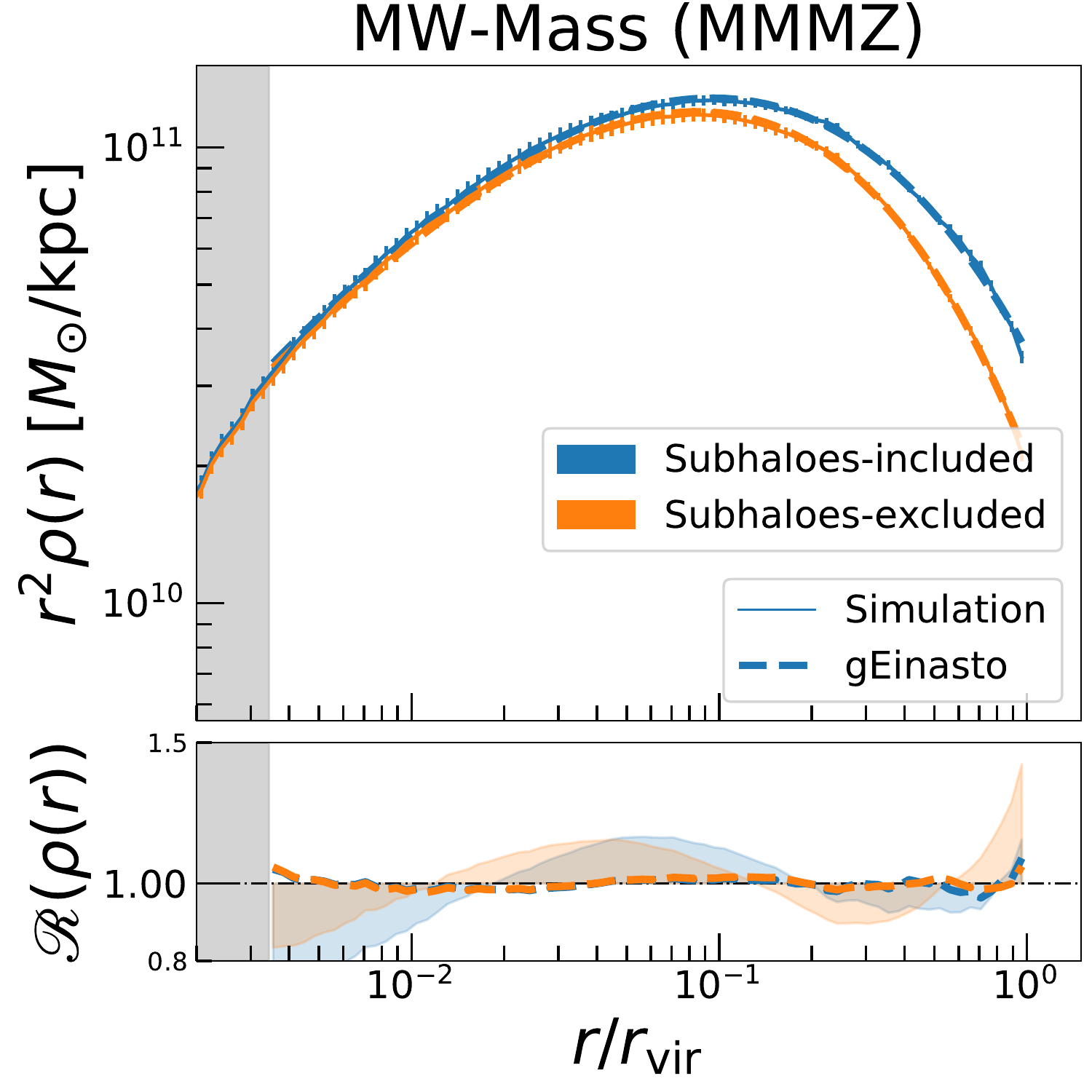}
        \caption{}
        \label{subfig:fullstack_mmmz}
    \end{subfigure}%
    \begin{subfigure}{0.5\textwidth}
        \includegraphics[width=0.9\textwidth]{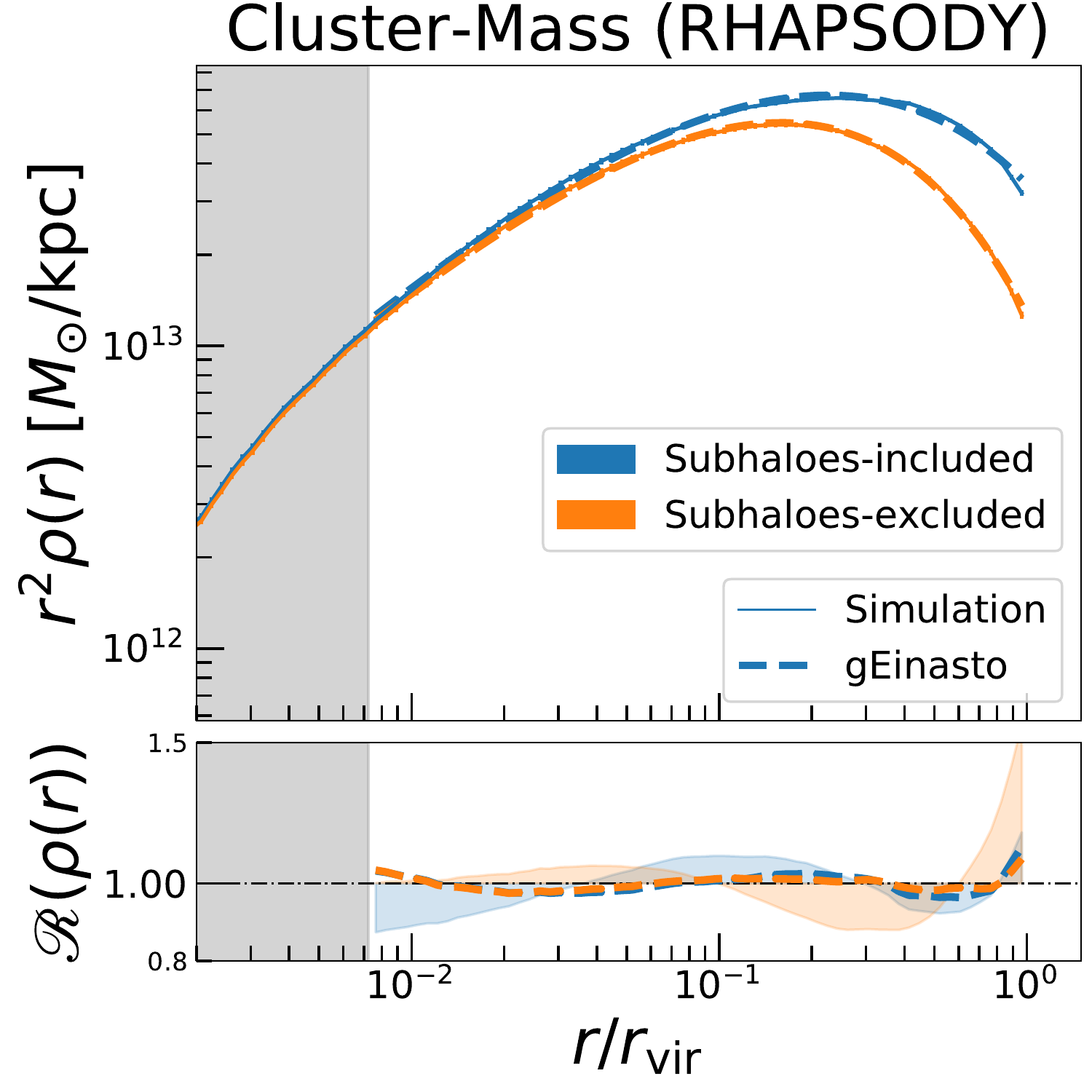}
        \caption{}
        \label{subfig:fullstack_rhapsody}
    \end{subfigure}%
\caption{\textbf{(a)} \textit{Upper panel:} The stacked density profile from all 45 Milky Way-mass (MMMZ) host haloes. The grey band at low $r$ indicates the adopted resolution limit of four times the softening length. The blue curve depicts the density profile when subhalo mass is included, as conventionally done in $N$-body analyses, while the orange curve corresponds to the case where subhaloes have been excluded. Both curves have been fit to a generalised Einasto profile, as defined by \autoref{eq:geinasto} and discussed in \autoref{Subsection:methods}. The thick blue dashed line is a generalised Einasto fit to the blue curve, and the thick orange dashed line is a fit to the orange curve. Although small, the plotted error bars correspond to the standard error, as describe in \autoref{subsection:fitting}. \textit{Lower panel:} Ratio of the density profile of the fit to the simulation data. The horizontal line at $\mathscr{R}(\rho(r)) = 1$ denotes where a perfect fit would lie. The blue and orange bands represent the area between the curve for the stacked profile fit to the standard NFW (described in \autoref{eq:nfw}) for subhalo-included and excluded respectively. In comparison, the dashed lines are much closer to the $\mathscr{R}(\rho(r)) = 1$ mark. It is evident that a generalised Einasto profile is a better fit to the simulation than the standard NFW profile in both cases as expected from \autoref{tab:stats}. \textbf{(b)} As in panel (a), but for the stacked density profile of the 96 cluster-mass (RHAPSODY) host haloes. In both mass ranges it is apparent that mass in subhaloes has the largest effect on the density profiles in their outer portions.} 
\end{figure*}

We now turn our attention to the optimal model, the generalised Einasto profile, and the comparison model, the NFW profile. \autoref{subfig:fullstack_mmmz} and \autoref{subfig:fullstack_rhapsody} show the stacked fits in both mass regimes.
As before blue denotes subhalo-included and 
orange denotes subhalo-excluded. The vertical 
grey band indicates radii below the simulation resolution limit 
(see \autoref{Subsection:simulations}). The fit to the 
generalised Einasto profile is shown by the thick dashed lines, and error bars represent the standard error in the stacked 
profile for each bin. These are among the same fits used 
to generate \autoref{tab:stats}.

The lower panels of \autoref{subfig:fullstack_mmmz} and \autoref{subfig:fullstack_rhapsody}
depict the ratio of the density profile fits to the simulated 
profiles. The $x$-axes align with the top panels while the $y$-axes show $\mathscr{R}(\rho(r))=\rho(r)_{\rm fit}/\rho(r)_{\rm simulation}$. This ratio would lie along the black dash-dot horizontal line at $\mathscr{R}(\rho(r)) = 1$ for a perfect fit; the smaller the deviation from the $\mathscr{R}(\rho(r)) = 1$ line, the closer a functional form is to the stacked profile. As in the upper panel the dashed lines are the fits to the generalised Einasto profile. The shaded bands show the regions between the $\mathscr{R}(\rho(r)) = 1$ line and $\mathscr{R}(\rho(r))$ values for NFW fits to the halo density profiles, for comparison. 

In terms of deciding upon the functional form that most 
faithfully represents simulated dark matter 
haloes, \autoref{tab:stats}, \autoref{subfig:fullstack_mmmz}, and \autoref{subfig:fullstack_rhapsody} all show that
at both mass ranges we have the same result: the gEinasto profile is a better representation of the simulation stacks than the NFW profile. The gEinasto profile has the possible disadvantage of two additional free parameters, 
but this result holds even using metrics, such as the 
BIC and AIC, that account for additional parameter 
freedom. These conclusions hold for both the subhalo-included 
and the subhalo-excluded data sets. 

Despite being described faithfully by the same functional form 
(the gEinasto profile), it is apparent that the subhalo-included and subhalo-excluded profiles are  notably different. In particular, subhalo-excluded density profiles drop more steeply at large radii than the subhalo-included density profiles. These differences 
manifest themselves in different best-fit profile parameters. The mass deficit at large halo-centric distance are a result of the fact that subhaloes are preferentially found in the outer regions of haloes \citep{zentner2005}. It is likely the case that the shallow outer slope of the NFW profile is a result of subhalo contribution, which we explore further in the following subsection. Without subhaloes, the host halos likely have higher concentrations as a result of this deficit of mass in the outer region, which we explore further in \autoref{subsection:c-m}.

\subsubsection{Effective Power Law Index}
\label{subsection:derivs}

One of the clearest ways to see \emph{why} the NFW (and other functional forms) do not perform as well as the gEinasto profile is by comparing the local power-law 
indices, defined as ${\rm d}\ln \rho/{\rm d}\ln r$, of the simulated profiles to various functional forms. 
Deviations from these asymptotic behaviours imply limitations to the quality of the fit that the corresponding profile can achieve. The technique for calculating the density derivatives is described in \autoref{Subsection:methods}.

\begin{figure*}
    \centering
    \begin{subfigure}{0.5\textwidth}
        \includegraphics[width=0.9\textwidth]{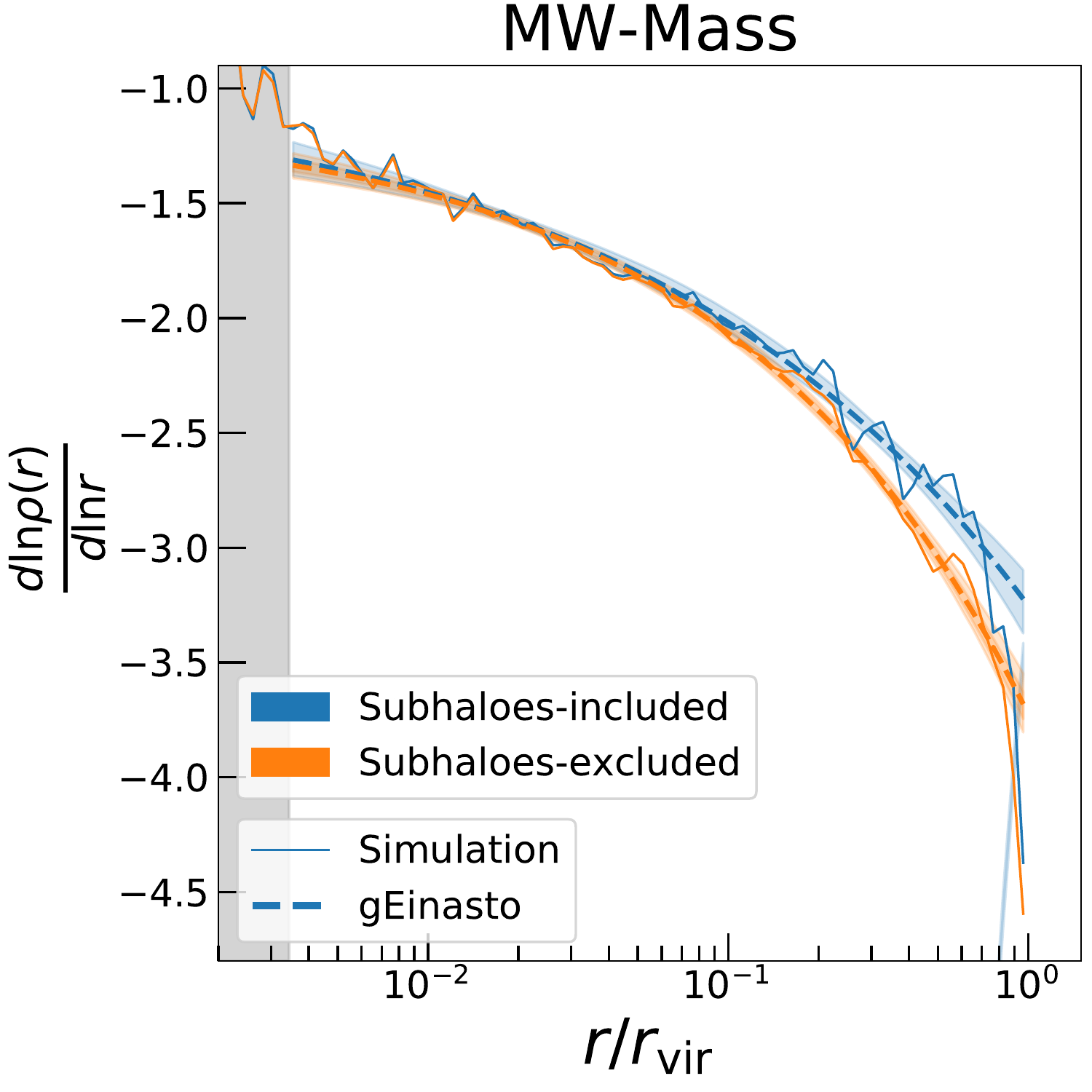}
        \caption{}
        \label{subfig:deriv_fiducial_mmmz}
    \end{subfigure}%
    \begin{subfigure}{0.5\textwidth}
        \includegraphics[width=0.9\textwidth]{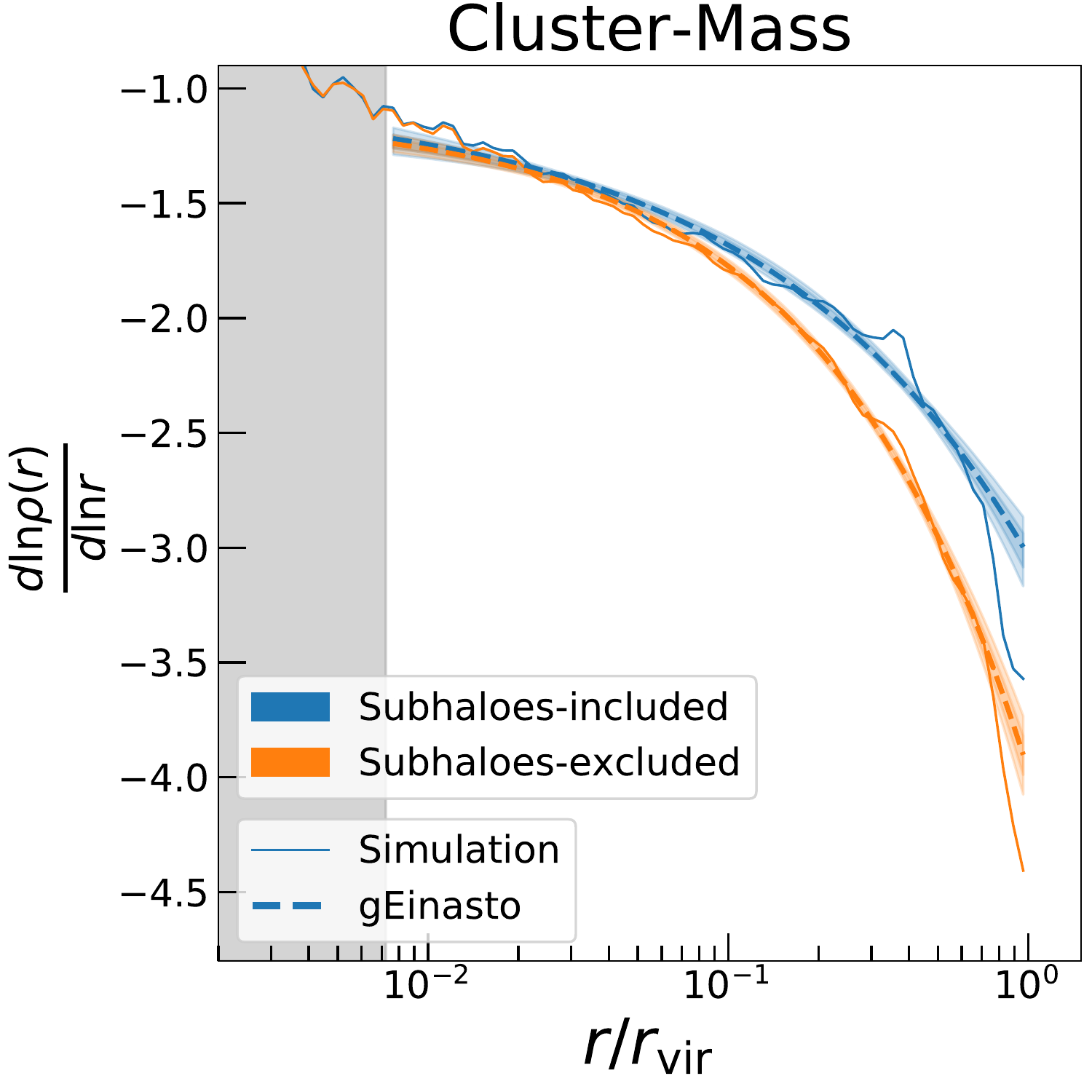}
        \caption{}
        \label{subfig:deriv_fiducial_rhapsody}
    \end{subfigure}%
\caption{\textbf{(a)} Effective power law index of the density profiles (i.e., logarithmic derivatives of the profiles) as a function of radius for the Milky Way-mass (MMMZ) haloes. Plotted here are the derivatives derived from a stack, or average, of the profiles from all of the MMMZ host haloes. The solid lines depict the derivative of the stacked simulation data, which is calculated using a 3-point derivative algorithm numerically. The dashed line shows the local power-law index implied from the fit to the generalised Einasto profile. We re-sample the 45 hosts with replacement and re-calculate the fits of these new samples to the generalised Einasto profile. Then we calculate the 68 and 95 percent regions of the local power-law index. The shaded blue and orange regions around the fit represent these bootstrapped errors. As before, the grey region represents radii below the resolution limit of the simulations. The generalised Einasto profile provides a good fit to the simulations except the very extreme outer region where the profile falls off. \textbf{(b)} Same as (a) but for the cluster-mass (RHAPSODY) haloes; the generalised Einasto profile is also a good descriptor of the simulations in this mass range.%
} 
\end{figure*}

In \autoref{subfig:deriv_fiducial_mmmz} and \autoref{subfig:deriv_fiducial_rhapsody}, we show the local power-law indices corresponding to the profiles shown in \autoref{subfig:fullstack_mmmz} and \autoref{subfig:fullstack_rhapsody}. The results shown in \autoref{subfig:deriv_fiducial_mmmz} and \autoref{subfig:deriv_fiducial_rhapsody} are \emph{not} 
new fits to the local power-law indices of the profiles. 
Rather, they show the local power-law indices implied by 
the fits to local density discussed above. In both plots the thin solid lines depict the local power-law indices of the simulation data and the thick dashed lines represent the parameters resulting from the fit to the generalised Einasto profile. 
The shaded regions around the fits to the gEinasto profile 
define the 68 and 95 percentile confidence regions of the fit. These percentile regions are calculated from 
bootstrap re-sampling (re-sampling with replacement) of the set of host halo density profiles 1000 times. For each bootstrap sample, the fit of the stack is performed again and the corresponding power law indices are computed by substitution of the fit parameters. Then we determine the confidence regions.

\autoref{subfig:deriv_fiducial_mmmz} and \autoref{subfig:deriv_fiducial_rhapsody} further demostrate that the generalised Einasto profile is able to capture the behaviour of the simulation for both the subhalo-included and the subhalo-excluded models.
It is also notable that the subhalo-excluded profiles are steeper than their counterparts due to the mass deficit in the outer region of the haloes.

\begin{figure*}
    \centering
    \begin{subfigure}{0.9\textwidth}
        \includegraphics[width=\textwidth]{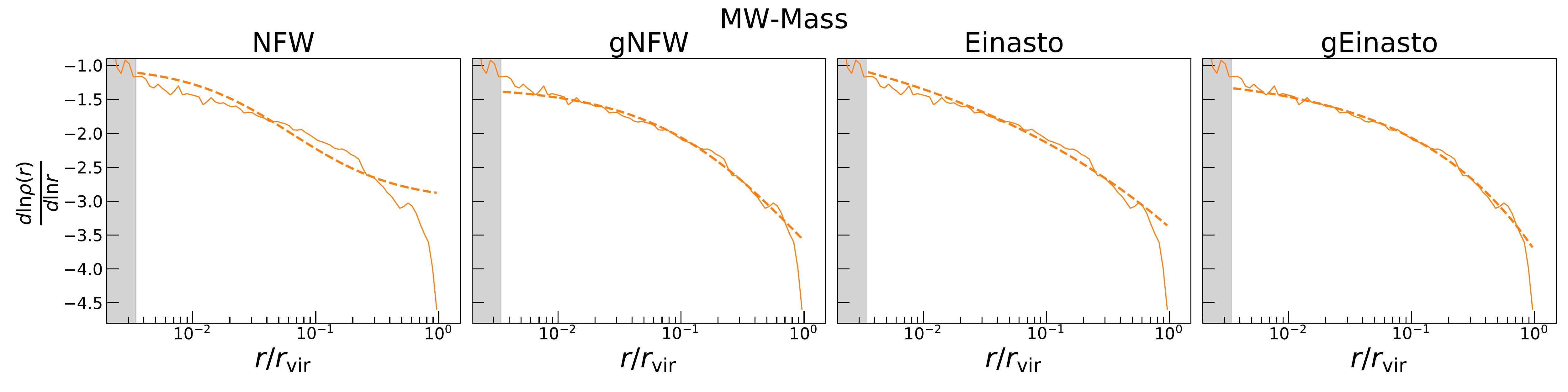}
        \caption{}
        \label{subfig:deriv_all_mmmz}
    \end{subfigure}%
    \\
    \begin{subfigure}{0.9\textwidth}
        \includegraphics[width=\textwidth]{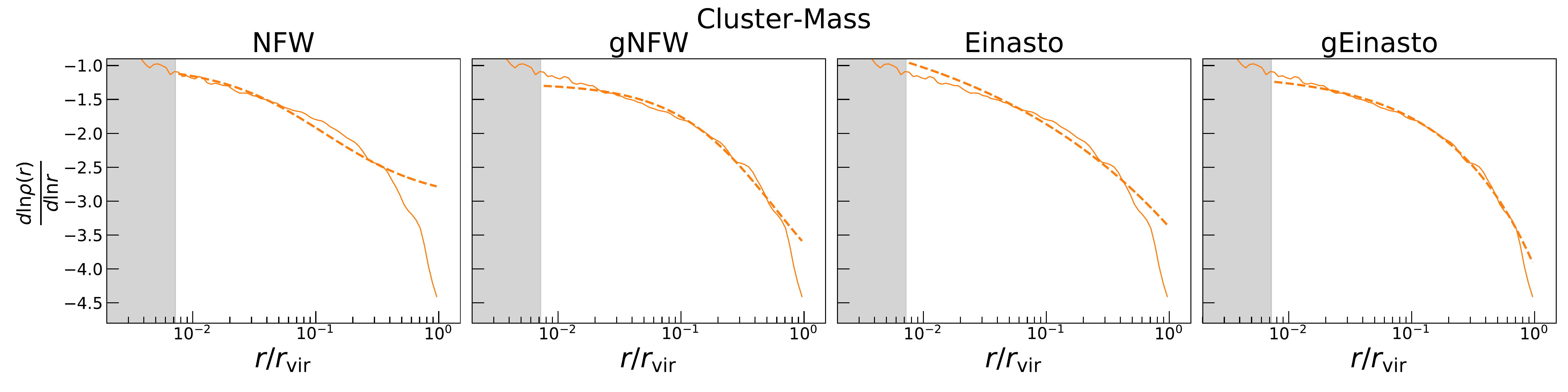}
        \caption{}
        \label{subfig:deriv_all_rhapsody}
    \end{subfigure}%
\caption{\textbf{(a)} Similar to \autoref{subfig:deriv_fiducial_mmmz}, this figure shows the effective power law indices of the density profiles as a function of radius for the MW-mass haloes for all four profiles (NFW profile, generalised NFW profile, Einasto profile, and generalised Einasto profile) explored herein. We show results for the subhalo-excluded models only, as all profiles provide good descriptions of the subhalo-included simulation data set. The solid lines depict the derivative of the stacked simulation data, and the dashed line shows the power-law index implied from the fit of the simulation data to the respective profile. \textbf{(b)} Same as (a) but for the cluster-mass haloes. For both mass regimes the NFW profile has a much shallower outer slope than the simulation data. The Einasto profile has a much less severe but shallow prediction as well. This shallow slope means the profiles do not have enough flexibility to match the simulation data when subhaloes are excluded. It is evident that the additional parameter of the generalised Einasto profile over the standard Einasto profile allows for a larger flexibility in describing the halo density profiles.}
\end{figure*}

We compare the subhalo-excluded power-law indices for all four profiles of interest (NFW, generalised NFW, Einasto, and generalised Einasto) in \autoref{subfig:deriv_all_mmmz} and \autoref{subfig:deriv_all_rhapsody}. The steep outer slopes of the subhalo-excluded simulation data set cannot be well described by either the NFW profile or the Einasto profiles. 

For the Einasto profile, the inner power-law index approaches 0 when $r \ll r_{\rm s}$. The parameter $\alpha$ allows the profile to steepen as $r$ increases, but no single value of $\alpha$ can capture the rate of increase of the slope on all scales. The gEinasto profile improves upon the Einasto profile by introducing a distinct parameter to capture the inner profile power-law 
index, making it approach to $-\gamma$ for $r \ll r_{\rm s}$. With this additional freedom, $\alpha$ in the gEinasto profile can be tuned to match the halo density profiles at $r \gtrsim r_{\rm s}$. In this way, the generalised Einasto profile is able to capture the shallowness of the subhalo-included profiles and the steepness of the subhalo-excluded profiles at large radii. 

The contrast between the outer power-law indices of the subhalo-included and subhalo-excluded profiles has a profound implication: the outer power-law index of the mass distribution of haloes is determined largely by subhaloes. A comparison between the results shown in \autoref{subfig:deriv_all_mmmz} and \autoref{subfig:deriv_all_rhapsody} to the general profile shown in \autoref{fig:gen_deriv} shows that the asymptotic behaviour of the given profile at $r \sim r_{\rm vir}$ drives the fit more than the asymptotic behaviour at the inner region. Profiles with relatively shallow slope at large radii, such as the standard NFW profile with ${\rm d}\ln \rho/{\rm d}\ln r = -3$, do not faithfully describe the smooth (subhalo-excluded) components of halos. This should be considered in any application for which one must model the smooth component of the host halo and the subhaloes associated with the host halo independently. 

\subsection{Impact of Subhaloes on Individual Halo Profiles}
\label{subsection:individ}

So far we have found, using stacked profiles and fit quality statistics, that the generalised Einasto profile describes well the halo mass distribution both with or without the presence of the mass in subhaloes. 
Haloes are dynamically evolving systems, which results in halo-to-halo variation for haloes of the same mass. A functional form that fits well to a stack may not necessarily be a good fit to the individual haloes that contributed to the stack due to this variation. In this subsection, we investigate fits to the profiles of individual simulated dark matter haloes, rather than to stacked profiles. We apply the same fitting procedures described in \autoref{subsection:fitting}.

\begin{figure*}
    \centering
    \begin{subfigure}{0.9\textwidth}
        \includegraphics[width=\textwidth]{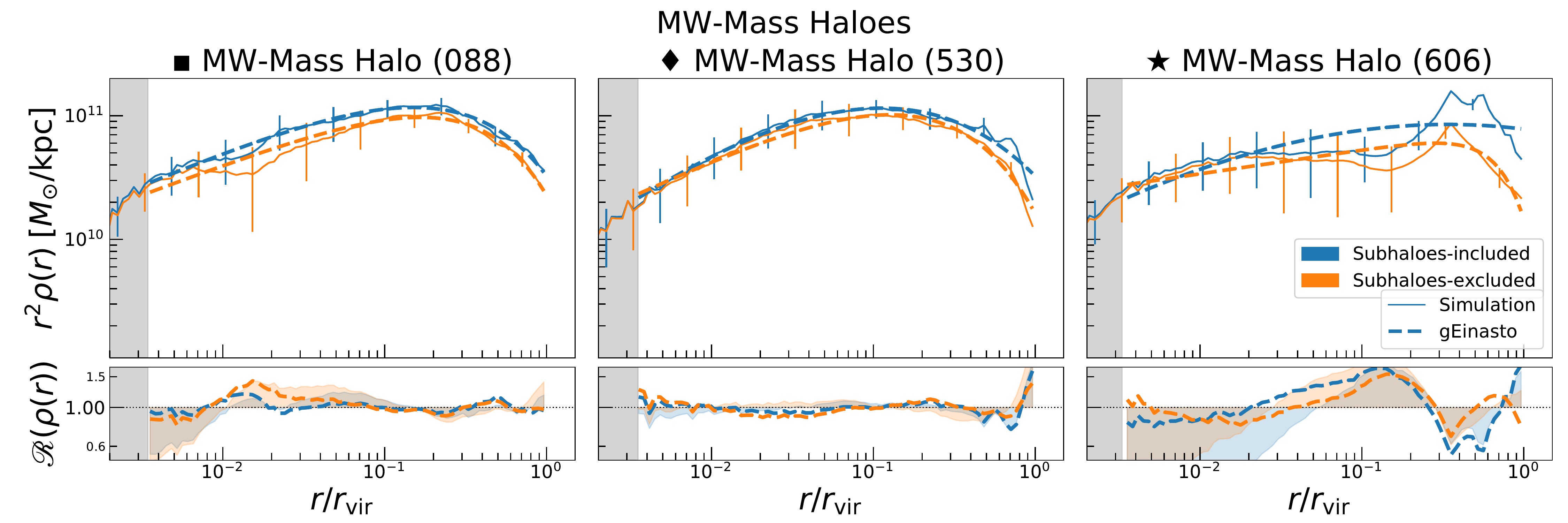}
        \caption{}
        \label{subfig:individual_mmmz}
    \end{subfigure}%
    \\
    \begin{subfigure}{0.9\textwidth}
        \includegraphics[width=\textwidth]{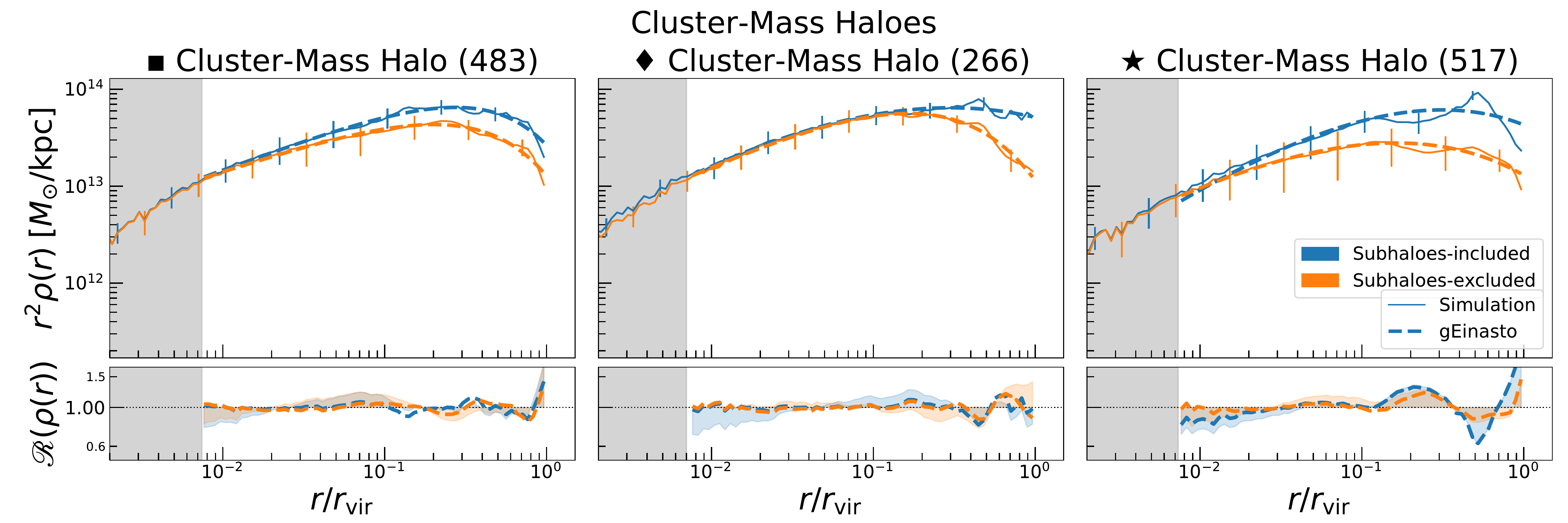}
        \caption{}
        \label{subfig:individ_rhapsody}
    \end{subfigure}%
\caption{\textbf{(a)} Three individual halo profiles and their respective fits from the MW-mass simulations (the axes and colours are the same as those in \autoref{subfig:fullstack_mmmz}). These haloes were selected according to the $\chisq$ of their fits \textit{including} subhaloes.  We have marked three haloes that are at the $33^{\rm rd}$ (Halo 088), $66^{\rm th}$ (Halo 530), and $99^{\rm th}$ (Halo 606) percentiles in $\chisq$, representative of a good, average, and not very good fit respectively. Error bars are representative of the halo-to-halo scatter, plotted every 10 bins starting at either the $0^{\rm th}$ bin (for subhalo-included; blue) or the $5^{\rm th}$ bin (for subhalo-excluded; orange). According to the residuals there is strong evidence that individual haloes tend to be better described by the generalised Einasto profile than the NFW profile. \textbf{(b)} As in (a) but for the cluster-mass host haloes. The haloes at the $33^{\rm rd}$, $66^{\rm th}$, and $99^{\rm th}$ percentiles in $\chi^{2}$ before subhalo removal are  Halo 483, Halo 266, and Halo 517. As in (a), the individual halo profiles are better fit to the generalised Einasto profile than the NFW profile.}
\end{figure*}

For the sake of brevity we show a selection of 3 haloes from each simulation in \autoref{subfig:individual_mmmz} and \autoref{subfig:individ_rhapsody}. We select the haloes closest to the $33^{\rm rd}$, $66^{\rm th}$, and $99^{\rm th}$ percentiles in \chisq{} from the subhalo-\textit{included} fits. This corresponds to Halo 088 (483), Halo 530 (266), and Halo 606 (517) for the MMMZ (RHAPSODY) haloes. We used the same fitting procedures as \autoref{subfig:fullstack_mmmz} and \autoref{subfig:fullstack_rhapsody} in axes and colours, the only difference being that the profiles are not stacked. The error-bars are representative of the standard deviation of all haloes in the respective simulation (the halo-to-halo scatter) of which the calculation is describe in (iv) of \autoref{subsection:fitting}.

Some of these profiles have notable peaks in the outer region or troughs in the inner region. We investigated these deviations from the fits for these haloes. These features are caused by massive subhaloes and/or active mergers. Consequently, these features are not well described by a single monotonic profile function when subhaloes are included, and in some cases even when subhaloes are excluded (as in Halo 088), because the particles are mixed in.

Despite the noisiness of the individual halo profiles, these results are very similar to those in \autoref{subfig:fullstack_mmmz} and \autoref{subfig:fullstack_rhapsody}. According to the residuals the generalised Einasto profile provides a tighter fit than the NFW profile. Additionally, as before, the subhalo-excluded profiles are much smoother. While the shapes of the profiles are not completely identical they are much more similar to each other.

\begin{figure}
    \centering
    \includegraphics[width=\linewidth]{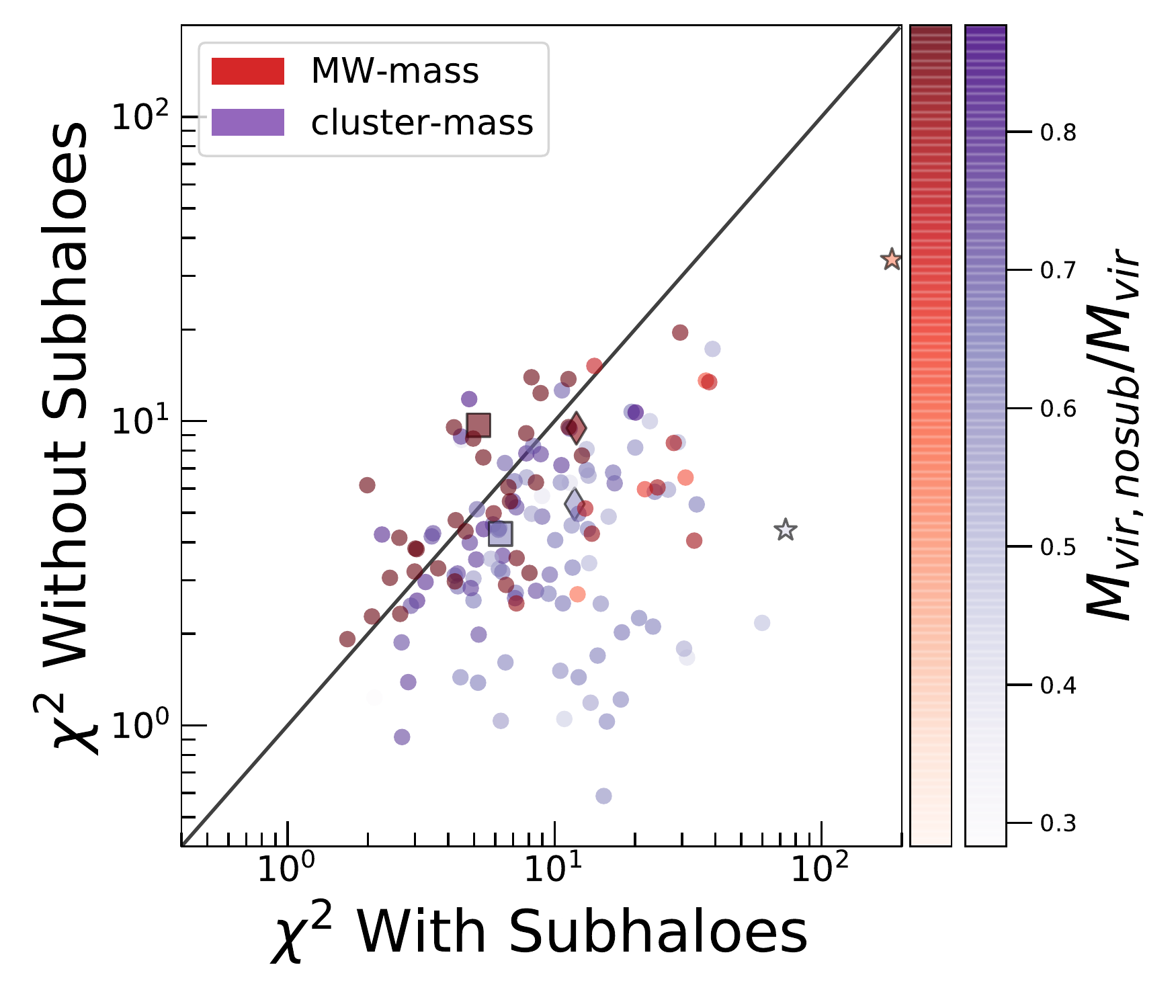}
    \caption{The \chisq{} of the individual halo fits to the generalised Einasto profile with subhaloes (the x-axis) plotted against the \chisq{} of the individual halo fits without subhaloes (the y-axis). Red points mark MW-mass (MMMZ) haloes and purple points mark cluster-mass (RHAPSODY) haloes. For reference, the black line across the diagonal shows $\chisq = \chisq_{\rm{nosub}}$. The three points marked here by the square, diamond, and star correspond to the halo profiles shown in \autoref{subfig:individual_mmmz} and \autoref{subfig:individ_rhapsody}. All of the points are coloured according to a proxy for subhalo number, the halo mass fraction in subhaloes $M_{\rm{vir, nosub}}/M_{\rm{vir}}$. The darker the shade, the less total halo mass in subhaloes (or the fewer subhaloes a halo has). The MMMZ haloes have quite a bit of scatter but tend to have improved \chisq{} after subhaloes are excluded. Nearly all of the RHAPSODY haloes have improved fits after subhaloes are excluded. In general the haloes with a larger mass fraction in subhaloes have smaller \chisq after subhaloes are excluded from the fits. Without subhaloes the halo profiles are smoother and have smaller residuals compared to analytic profiles. However, the scatter indicates that subhalo abundance alone does not account for the change in \chisq{}.}
\label{fig:chi2}
\end{figure}

To assess fit quality, \autoref{fig:chi2} shows the \chisq{} of the individual halo fits to the generalised Einasto profile, where the $x$-axis is the \chisq{} with subhaloes and the $y$-axis is the \chisq{} without subhaloes. Red points mark MMMZ haloes and purple points mark RHAPSODY haloes. The black, diagonal line corresponds to the case in which both values of \chisq{} are equal. The square, diamond, and star points correspond to the halos whose profiles are shown in \autoref{subfig:individual_mmmz} and \autoref{subfig:individ_rhapsody}.

It is evident that the \chisq{} values for the fits to the gEinasto profiles are generally smaller when subhaloes are excluded. This is the case for both simulation mass ranges, but the effect is more pronounced in the RHAPSODY haloes. This decrease in \chisq{} is likely a result of the halo density profiles yielding a much smoother mass distribution after subhaloes are removed. Because we are fitting to a smooth functional form, this form can describe the halo density profile with smaller residuals.

We further investigate if this change in \chisq{} is correlated with the fraction of mass in subhaloes, defined as $M_{\rm{vir, nosubs}}/M_{\rm vir}$, where $M_{\rm{vir, nosubs}}$ is the total mass within $R_{\rm vir}$ but excluding particles associated with subhaloes. The points in \autoref{fig:chi2} are shaded according to the subhalo abundance metric; halos marked in darker colours have a smaller mass fraction in subhaloes. The haloes with a greater portion of their mass in subhaloes tend to have worse \chisq{} when subhaloes are included, an effect which is substantially alleviated without subhaloes. This follows the narrative that the halo density profiles are much smoother without subhaloes and thus better described by an analytic profile. However, there is a decent scatter about this trend - the haloes with the most mass in subhaloes don't necessarily have the worst fits to the generalised Einasto profile. Thus subhaloes abundance alone cannot describe this change in \chisq{}.

Overall we have shown that in addition to stacked haloes individual halo profiles can be well described by the generalised Einasto profile. Both by eye inspection (examining the profile fits) and the results of the \chisq{} of the fits provide strong evidence that the generalised Einasto fits to the individual profiles are acceptable. We also show that regardless of stacks or individual profiles, halo density profiles exhibit a mass deficit in the outer halo region when subhaloes are excluded. Additionally the subhalo excluded component of the haloes are smoother than their subhalo included counterparts, which allows them to be better described by a smooth functional form.

\subsection{The Concentration--Mass Relation With and Without Subhaloes}
\label{subsection:c-m}

It is now well known that mean halo concentration is a slowly declining function of halo mass \citep[e.g.,][]{bullock2001,eke2001,wechsler2002,zhao2003,duffy2008,gao2008,maccio2008,klypin2016,ludlow2016,prada2012,diemer2015}. This mass dependence of halo concentration is widely thought to be caused by the fact that larger haloes assemble their masses later, on average, than their less-massive counterparts. In this section, we study the degree to which subhaloes themselves influence the mass dependence of halo concentrations. We emphasise that the concentrations calculated when subhaloes are excluded use the original virial radius of the halo with subhaloes. There is no unambiguous way to define a halo without its subhaloes so here we take the simplest approach of maintaining the same halo ``edge'' in both cases.

\begin{figure}
    \centering
    \includegraphics[width=\linewidth]{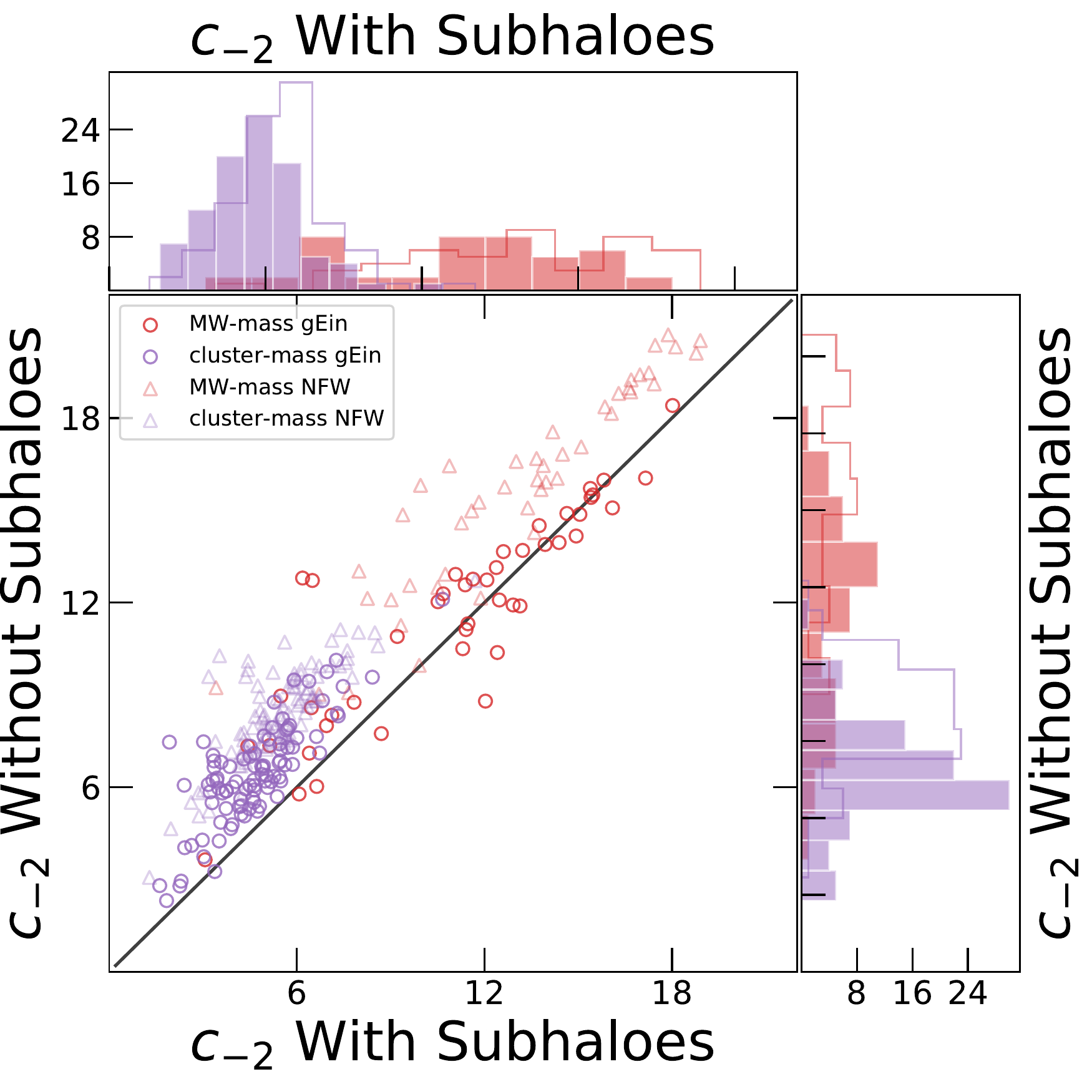}
\caption{The concentration of the MW-mass MMMZ haloes (red) and the cluster-mass RHAPSODY haloes (purple) calculated from individual fits to the generalised Einasto profile (open circles) and the NFW profile (open triangles) as described in \autoref{Subsection:methods}. The horizontal axis is the concentration when including the mass in subhaloes, denoted by $c_{-2}$ With Subhaloes, and the vertical axis is the concentration when excluding the mass in subhaloes as denoted by $c_{-2}$ Without Subhaloes. We also show corresponding histograms of the concentrations for the gEinasto profile (filled) and the NFW profile (outlined). A solid black line shows $y=x$; points on this line result in the same concentration values regardless of the presense of subhaloes. The MMMZ haloes show little to no change in gEisnasto concentration after excluding subhalo mass, while the RHAPSODY haloes almost all have notably higher gEinasto concentrations without subhaloes. For NFW concentrations, there is a larger increase in both mass ranges when halos are excluded. We find that a concentration--mass relation remains, even with the exclusion of subhalo mass.}
\label{fig:concentration}
\end{figure}

The best-fit values of halo concentration for subhalo-included and -excluded are shown in \autoref{fig:concentration}. Red points denote the MMMZ haloes and purple points denote the RHAPSODY haloes. Open circles mark values calculated from the generalised Einasto fits and open triangles mark values calculated from the NFW fits. We also show corresponding histograms of the concentration values, filled histograms represent the generalised Einasto profile results and outlined histograms represent NFW profile results. The concentration parameter is calculated as described in \autoref{Subsection:methods}. 

It is evident that haloes of different mass ranges exhibit different changes in concentration. The MMMZ haloes roughly maintain the same concentrations, with or without subhaloes, from the generalised Einasto fits. However these same haloes have increased concentrations resulting from the NFW fits. In contrast, the RHAPSODY haloes almost all have higher concentrations without subhaloes regardless of the fit profile. In part this effect is a result of the more massive RHAPSODY haloes having a larger fraction of their mass in subhaloes. However, despite this change the concentrations of the RHAPSODY haloes and MMMZ haloes do not match without subhaloes. This implies that the concentration dependence on halo mass is not purely decided by the presence of subhaloes --- the smooth central halo component is also impacted by halo formation history.

In \autoref{tab:c2} we quantify the changes exhibited by the concentration after subhaloes are excluded, with similar conclusions to those described above. The errors presented are calculated via bootstraps of the haloes. In the first and third columns we present the median concentrations of the individual halo fits (for MMMZ and RHAPSODY haloes respectively). The generalised Einasto profile median concentration has a smaller increase without subhaloes compared to the NFW profile (and even decreases minimally for the MMMZ haloes). Because subhaloes are typically in the outskirts of haloes, the exclusion of the mass in subhaloes results in an increase in concentration \citep{zentner2005}. 

Next we calculate the scatter of the concentration at a fixed halo mass, $\sigma_{\log c_{-2}}$, shown in columns two and four of \autoref{tab:c2}. This is done by calculating the interquartile range (IQR) of $\log c_{-2}$. This gives an estimate of $\sigma$, where we assume $\log c_{-2}$ follows a normal distribution such that $\sigma = \rm{IQR}/1.349$. When subhaloes are excluded, the scatter in both simulations is much smaller, indicating less scatter amongst halo fitting parameters. In general the scatter is smaller in the more massive RHAPSODY haloes compared to the MMMZ haloes, which is an expected result from, e.g., \citet{neto2007,duffy2008}. The subhalo-included results fall within the range of values estimated for the NFW profiles \citep[see e.g.,][]{jing2000,bullock2001,wechsler2002,comerford2007,neto2007,duffy2008}. 

The final column of \autoref{tab:c2} shows the ratio of the 
median concentration of the Milky Way-mass MMMZ haloes to the 
median concentration of the cluster-sized RHAPSODY haloes, $\frac{\mathrm{MMMZ}\;\mathrm{med}(c_{-2})}{\mathrm{RHAPSODY}\;\mathrm{med}(c_{-2})}$. This ratio is larger than one in all cases, reflecting the fact that concentration is a slowly decreasing function of halo mass. We find that the value of this ratio computed from the 
{\em subhaloes-excluded} concentrations is smaller than the value for the standard, {\em subhaloes-included} concentrations. The median concentrations of the haloes of the different masses exhibit a higher degree of similarity in the {\em subhaloes-excluded} case. Halo concentrations still exhibit a non-negligible mass dependence when subhalos are excluded.  Thus subhaloes increase the mass dependence of concentrations but do not completely explain the mass trend.

\begin{table*}
\centering
\includegraphics[trim=10mm 0mm 0mm 0mm, clip,width=0.99\linewidth]{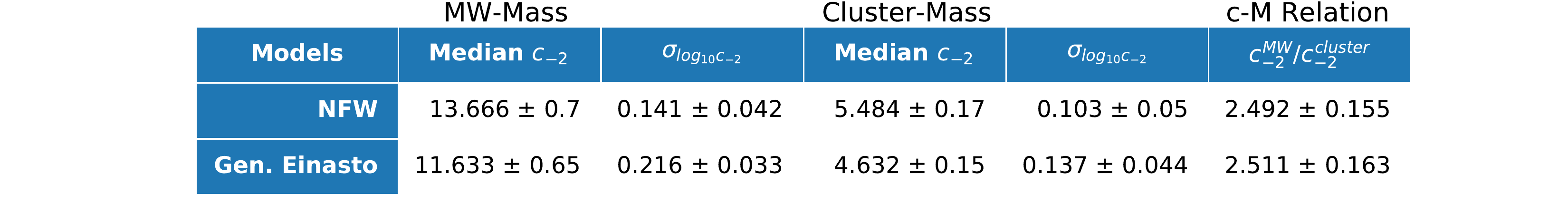}
\includegraphics[trim=10mm 0mm 0mm 0mm, clip,width=0.99\linewidth]{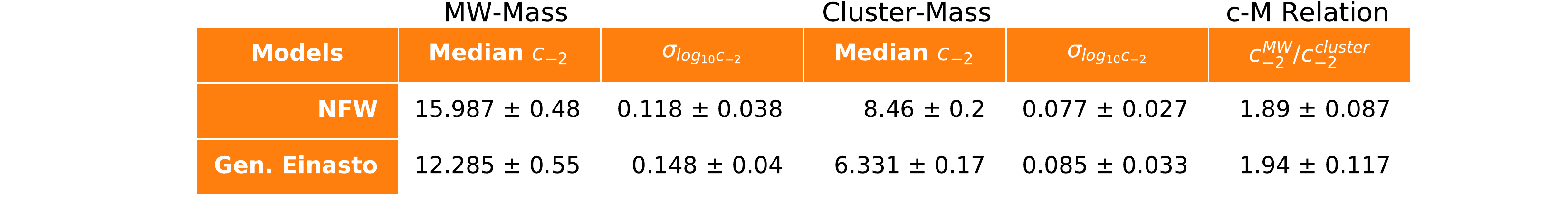}
\caption{Tables of the median concentrations and scatter for the simulated haloes. The concentrations are calculated as described in \autoref{Subsection:methods}. As before the blue table shows results of subhalo-included models and the orange table shows results for subhalo-excluded models. The median concentrations are simply the medians of the values computed for each individual halo for the given model, with errors from bootstrapping. We also list concentrations of the stacked halo profiles in \autoref{tab:best_fit}. The scatter $\sigma_{\log c_{-2}}$ describes the scatter in the concentration--mass relation. Concentration is simply a transformation of the halo parameters and makes it easy to study the differences, namely that the scatter is much smaller for subhalo-excluded models indicating that their parameters are more similar. The final column $c_{-2}^{\rm MW}/c_{-2}^{\rm cluster}$ is the fraction of the median concentrations for the respective simulation. The fractional change in concentration is much smaller for subhalo-excluded. These results indicate that subhaloes have an effect on the concentration-mass relation but do not completely explain the trend.}
\label{tab:c2}
\end{table*}

\subsection{Robustness Checks and Caveats}
\label{subsection:consistency}

We have performed several tests to ensure the robustness of our results. First, we examined the results when stacking subsets of haloes according to additional halo properties to see if this had an effect on the fit assessment (such as subhalo abundance). We have tabulated the same statistics as those in \autoref{tab:stats} when stacking subsets of the haloes split according to their subhalo mass fraction (see \autoref{subsection:individ} for further discussion on this quantity) and other such proxies for subhalo abundance such as metrics for halo formation time (as halo formation time is anti-correlated with subhalo count). In general the trends were the same as those shown for the full stack, barring a couple of edge cases where the stack was influenced by a few haloes undergoing mergers. 

In general, we present fRMS values calculated using the logarithmically spaced bins defined in (i) of \autoref{subsection:fitting}. Logarithmic bins have the effect of preferentially weighting the inner portions of halo profiles relative to a binning scheme linear in radius, $r$. To do that we computed a linear fRMS by interpolating densities onto a linear grid and recomputing each of the fits. We have confirmed that using a linear binning scheme does not alter the qualitative results and preserves the rank ordering of fRMS among haloes in the sample.

A separate concern is that the halo profiles have not turned over yet at the virial radius, especially for the more massive RHAPSODY haloes. The edge of a halo is arbitrarily defined. Because of this there is the possibility that the virial radius may no be the physical edge of the halo. In terms of density profiles this means that the density is not yet at the point where the slope is diverging. We have examined and fit profiles extending to both $1.5\times r_{\rm vir}$ and $2\times r_{\rm vir}$ and find that this has no effect on our qualitative results.

In addition to these consistency checks, 
our results and conclusions are subject to several caveats. First, concentrations were determined via a specific algorithm for halo profile fitting. While we expect the qualitative aspects of our conclusions to be unaltered by the application of distinct algorithms, some of the quantitative details of the results may be sensitive to the algorithm used to determine halo concentrations. 

Second, the detailed results presented here are dependent on what one defines as ``substructure.'' In \autoref{subsection:calc} we define streams, caustics, and loosely self-bound objects that are actively being disrupted as part of the smooth halo component. Including these objects in the substructure component would reduce the mass of the smooth halo component and may also systematically alter the 
structure of the smooth halo component. 
For further detail and examples of an alternative way of defining substructure see Appendix~\ref{Section:rockstar_sub_removal} and relevant figures. Even with that alternate definition, the generalised Einasto profile still has the smallest BIC, fRMS, \chisq, etc for subhalo excluded fits.

In this work we use the default binding criteria defined in \textsc{Rockstar} (see \citet{behroozi2013} for further detail), or the limit that a subhalo must meet in order to be self bound. Using different halo finders and different binding criteria can result in slightly different substructure abundances. This in turn can slightly alter what is counted as ``substructure'' and what is attributed to the smooth halo component. While we use the criteria that $50\%$ of particles must be bound to a subhalo, \citet{behroozi2013} finds that effects don't manifest until a threshold of $15\%$ or lower. We expect our qualitative results to hold for different subhalo self-binding criteria, especially because the subhalo mass fraction is generally small compared to the overall halo mass.

Third, as with all simulation-based studies, the simulations that we analysed had finite force and mass resolution. We again expect that our qualitative conclusions are not compromised by finite resolution, but several quantitative aspects of the results may be resolution dependent. Most notably, we provide best-fit parameters for each profile in Appendix~\ref{Section:best_fit_vals}. It may be tempting to take the inner profile power-law indices ($\gamma$) to represent the asymptotic inner power-law indices of halo 
density profiles, as this is a quantity of interest across many sub-disciplines. However, we caution that the specific power-law indices quoted are 
likely resolution dependent and our analysis cannot 
to rule out profiles that become shallower than 
$\rho \propto r^{-1}$ at small radii. 

Resolution can alter our quantitative results in at least one additional way. As resolution increases, smaller subhaloes will be resolved within the simulation. Consequently, removing ``subhaloes'' according to the definition used herein is inherently resolution dependent. In future studies, the most sensible definition of {\em subhalo} and/or {\em substructure} will likely depend upon the specific data analysis that the simulation results will be compared to.

Fourth, we have only examined two narrow halo mass bins here. Therefore, we cannot make specific statements about the detailed mass dependence of any of the effects that we have explored, 
including the concentration--mass relation. An interesting extension of this work would be to consider the profiles of haloes over a wider range of masses in order to construct an improved halo concentration--mass relation based on smooth profiles. 

Lastly, recent work such as that by \citet{carlsten2020} finds that satellites are more concentrated than subhaloes. Baryonic effects may alter our quantitative results along these lines; while in $N$-body simulations the subhalo mass loss occurs only in the outer regions, it may occur at smaller radii when baryonic physics is introduced. Another interesting extension of our work would be to explore the effects of removing subhaloes in hydro-dynamical models.

\section{Summary and Conclusion}
\label{Section:Conclusion}

In this paper, we have investigated the density profiles of the smooth components of host dark matter haloes and compared them with conventional halo density profiles. Typically, the density profiles of host dark matter haloes are analysed including all of the mass associated with subhaloes within these hosts. Here, we isolate the smooth components of the host haloes by removing the mass associated with subhaloes \citep[following earlier work by][]{wu2013}, and study the resultant smooth host density profile. 

We examine the difference between the smooth and 
conventional density profiles for a set of high-resolution simulations of 45 Milky Way-sized haloes 
(the MMMZ haloes, with $M_{\rm vir} = 10^{12.1\pm0.03}\msun$), 
and a set of simulations of 96 cluster-sized 
haloes (the RHAPSODY haloes, 
with $M_{\rm vir} = 10^{14.8\pm0.05}\msun$). 
Considering profiles at different masses is a priority 
because the amount of halo substructure is known to increase systematically with halo mass \citep{zentner2005}. However, the prerequisite of high resolution precludes an exploration of a wide range of halo masses. Studying high-resolution simulations 
of Milky Way- and cluster-sized haloes is a 
compromise between these considerations. 

We have drawn four primary conclusions from our 
work, which can be summarised as follows.
\begin{enumerate}[leftmargin=0.9\parindent,rightmargin=0.9\parindent]
    \item[{\bf (i)}] 
    The density profiles of the smooth components 
    (i.e., excluding mass within subhaloes) of host 
    dark matter haloes decline more steeply at large radii compared to conventional density profiles that include both smooth mass and mass within subhaloes.\\
 
    \item[{\bf (ii)}] A single functional form, the 
    generalised Einasto (gEinasto) profile, describes 
    all of the profiles that we have studied, including 
    both smooth (subhaloes-excluded) and conventional 
    (subhaloes-included) profiles, with a smaller residual error than either the often-used NFW or Einasto profiles.\\
    
    \item[{\bf (iii)}] 
    We find that concentrations ($c_{-2}$) derived 
    from the smooth halo density profiles exhibit a 
    weaker dependence upon mass than the concentrations 
    derived from conventional density profiles including subhaloes. This indicates that substructure
    plays an important role in establishing the 
    concentration--mass relation.\\
    
    \item[{\bf (iv)}] Concentrations derived 
    from the density profiles of the smoothed 
    components of haloes exhibit exhibit smaller 
    scatter at fixed mass than conventional 
    concentrations. This indicates that substructure plays a role in establishing the distribution of halo 
    concentrations at fixed mass.
\end{enumerate}
Each of these conclusions has a variety of important 
consequences.  We elaborate on 
each of points {\bf (i)}--{\bf (iv)} in turn below.

The prevalence of substructure is a natural consequence of CDM.  It is also known 
that subhaloes are distributed within their hosts differently than the mass distribution
\citep{nagai2005,zentner2005}.
One consequence of this difference is 
that the smooth component of a halo has a 
density profile that 
is different from its total mass density profile including subhaloes, declining more rapidly than the full halo density profile. This distinction may be relevant to studies that aim to understand the nature of the nearly universal 
density profiles of haloes, and may impact analyses of a variety of observations.  The effect that we 
measure is also broadly consistent with recent work 
describing the influence of mergers on halo 
concentrations \citep{wang2020}. 
 
Consider the analysis of a hypothetical gravitational lens system as an illustration of the importance of the distinction between the mass in the smooth component of the host and its subhaloes.  The mass distribution of the lens system can be constrained through the observation of the magnified/distorted images of the source 
galaxies behind the lens system. A common strategy is 
to treat the bulk of the lens mass as an NFW profile. 
However, visible substructures (satellite galaxies) or 
invisible substructure are treated as haloes with their own, distinct profiles. The problem with this scheme is that the NFW profile used to describe the main lens 
system is already calibrated to include the mass in substructure.  This new, composite lens system, 
built from an NFW main halo and distinct subhaloes, 
no longer represents the mass distributions 
found in CDM halo simulations. One might informally say that the subhaloes are ``double counted.'' The model would represent the predictions of simulations more faithfully if the main lens were modelled using a profile calibrated 
to the smooth component of the host halo alone. This work provides such profiles.
 
Similarly, halo density profiles determine the luminosities of extra-galactic sources of dark matter annihilation. The boost factors associated with the balance between the smooth halo component and subhaloes could be altered by the differences between the two as we have shown such profiles are markedly different.

We demonstrated that the ``generalised Einasto'' 
(gEinasto) profile introduced here provides a better description of  dark matter halo density profiles than either the NFW or Einasto profiles (or several other 
candidate profiles, see Appendix~\ref{Section:other_profiles}). 
Indeed, in all cases that we have studied, 
including MMMZ (Milky Way-sized) haloes and 
RHAPSODY (cluster-sized) haloes, both with and 
without including the mass within subhaloes 
in the density profiles, the gEinasto profile 
provides a superior description of the dependence 
of halo density on halocentric distance. 
As evinced in \autoref{fig:gen_deriv}, \autoref{subfig:deriv_all_mmmz}, and \autoref{subfig:deriv_all_rhapsody}, 
the Einasto profile has the shortcoming that 
it cannot describe the variation of the 
local power-law index 
${\rm d}\ln \rho/{\rm d}\ln r$ as a function of 
$r$ with a single value of $\alpha$ for the entire halo. 
The gEinasto profile is 
an Einasto-type profile with an additional 
free inner power-law index. This additional 
freedom enables the Einasto profile to describe 
halo density over the entire range of resolved 
halo radii. A gEinasto fit to stacked halo density 
profiles describes the haloes with residuals smaller 
than a few \% in all cases.

Concentration ($c_{-2}$) is a convenient dimensionless characterisation of the scale at which a halo profile steepens. Here we find that the 
concentrations of the smooth halo density profiles have a weaker dependence on mass than conventional concentrations, which is shown in \autoref{fig:concentration} and \autoref{tab:c2}. First, we find that the concentrations for the subhalo-excluded profiles (for the gEinasto profile) are up to $\sim30\%$ higher than concentrations derived from the conventional halo mass distribution for individual haloes (and up to $\sim35\%$ higher for the stacks). This is in agreement with results from past work \citep[e.g.,][]{zentner2005,mao2015,fielder2019} that indicated that haloes with higher concentrations tend to have fewer subhaloes. W also find higher halo concentrations when subhaloes are removed. However, a mass dependence on concentration remains, even without subhaloes in the picture. Thus the concentration--mass relation is only partially explained by subhaloes.

At fixed mass we also find that the smooth profile concentrations have a significantly smaller scatter than concentrations with subhalos included.  Some scatter remains after subhalo removal.  One possible source of this remaining scatter is environment, as work by \citet{maccio2007} found that more concentrated haloes live in denser environments at fixed mass. This is consistent with the now well-known concentration-dependent clustering of haloes \citep{wechsler2006,gao2007}. 

We have shown that subhaloes have a prominent effect on 
the profiles of dark matter haloes at larger radii 
($r \gtrsim r_{-2}$), and have given fitting formulae that encapsulate this effect. As we collect ever more precise 
data at a variety of observational facilities, 
we hope that this new accounting for halo mass will enable more powerful and less biased data analyses. The effort to understand the role of subhaloes in determining 
halo properties is only beginning. Given this work, it is 
reasonable to suspect that subhaloes may influence a variety of halo properties. While a detailed exploration is beyond the scope of this paper, future studies of such effects may yield tools which can further improve data analyses and deeper insights into the formation and evolution of dark matter haloes.

\section*{Acknowledgements}
The authors thank Peter Behroozi for explaining detailed features of \textsc{Rockstar}, and thank Oliver Hahn for work in the RHAPSODY simulations. 
The MMMZ and Rhapsody simulations used in this research were created with computational resources at SLAC National Accelerator Laboratory, a U.S.\ Department of Energy Office; Y.-Y.M., H.-Y.W., and R.H.W.\ thank the support of the SLAC computational team.
Support for Y.-Y.M.\ was provided by the Pittsburgh Particle Physics, Astrophysics and Cosmology Center through the Samuel P.\ Langley PITT PACC Postdoctoral Fellowship, and by NASA through the NASA Hubble Fellowship grant no.\ HST-HF2-51441.001 awarded by the Space Telescope Science Institute, which is operated by the Association of Universities for Research in Astronomy, Incorporated, under NASA contract NAS5-26555. The work of ARZ was supported, 
in part, by grant AST 1517563 from the US National Science Foundation (NSF). 

This research made use of Python, along with many community-developed or maintained software packages, including IPython \citep{ipython}, Jupyter (\http{jupyter.org}), Matplotlib \citep{matplotlib}, NumPy \citep{numpy}, Pandas \citep{pandas}, and SciPy \citep{scipy}.
This research made use of NASA's Astrophysics Data System for bibliographic information.

\bibliographystyle{mnras}
\bibliography{refs,software}

\clearpage
\appendix
\section{Other Halo Profiles}
\label{Section:other_profiles}
There have been numerous papers discussing various profiles that describe dark matter haloes. Most are variations of the double power-law generalised NFW profile (\autoref{eq:gennfw}) with power-law 
indices $\alpha$, $\beta$ and $\gamma$ set to various specific 
values. There has been more recent work that also studies varieties of continuously varying power laws, namely modifications to the Einasto profile (\autoref{eq:einasto}). In addition to the primary results 
presented in the main text, we investigated fitting dark matter halo 
density profiles to the following analytic forms.
\begin{enumerate}
        \item The generalized NFW profile with various constraints on $\alpha$, $\beta$, or $\gamma$. E.g., $(\alpha,\beta,\gamma) = (1,3,\gamma); (1,\beta,\gamma); (\alpha,\beta,1.58)$.
        \item The Generalised Moore profile \citep{moore1999} ($\alpha$, $\beta$, $\gamma$) = ($3-\gamma$, $3$, $\gamma$).
        \item The Denhen \& McLaughlin Profile \citep{denhen2005} ($\alpha$, $\beta$, $\gamma$) = ($\frac{4}{9}$, $\frac{31}{9}$, $\frac{7}{9}$).
        \item The Generalised Denhen \& McLaughlin Profile \citep{denhen2005} ($\alpha$, $\beta$, $\gamma$) = ($\frac{3-\gamma}{5}$, $\frac{18-\gamma}{5}$, $\gamma$).
        \item The \citep{dicintio2014} model ($\alpha$, $\beta$, $\gamma$) = ($0.84$, $2.85$, $1.09$) using the equations in their paper, and $M_{*} = 5 \times 10^{10}$ and $M_{\rm halo} = 1.3 \times 10^{12}$ for the Milky Way from \citet{bland2016}.
        \item A log parabola (or curved power law), as it was a good descriptor of the effective power law index of our profiles (i.e., the derivative of the log of the density with respect to log r, which is discussed in \autoref{subsection:derivs}). This is expressed by $\rho(r) = (\frac{r}{r_{s}})^{-\alpha-\beta \ln{(r/r_{s})}}$. 
\end{enumerate}
None of these profiles performed as well as the generalised NFW or generalised Einasto profiles. We mention them here for completeness for the reader. 

\section{Subhalo Removal with \textsc{Rockstar}}
\label{Section:rockstar_sub_removal}

Here we continue the discussion of how subhaloes are excluded from calculations. Cleanly identifying particles that belong to detectable subhaloes and distinguishing them from particles that belong only to the host and not to any subhalo is not straightforward. Objects identified as subhaloes in the halo table do not correspond one-to-one to the set of over-densities visible in particle distributions. For example, \textsc{Rockstar} does not list subhaloes with low self-bounded particle fractions in the halo catalogue ($<50\%$ self-bound particles). These are structures that are very diffuse and do not fall within the strict "subhalo" definition. Some of these objects are very small and not real, so one must define a mass and self-bound fraction cut in order to get high purity results. Please refer to the \textsc{Rockstar} paper, \citet{behroozi2013}, for details on binding criteria.

In \autoref{subsection:calc} our "subhalo-excluded" particles are those that are not associated with any subhalo listed in the \textsc{Rockstar} halo catalogue. This sample includes objects that do not meet the bounded-ness criteria in \textsc{Rockstar}. However, it is also possible to select particles that are \textit{only} tagged as host particles. This would mean excluding both particles that belong to substructure that meets \textsc{Rockstar}'s criteria \textit{and} particles that are part of other more diffuse substructures.

\begin{figure*}
    \centering
    \includegraphics[width=0.9\linewidth]{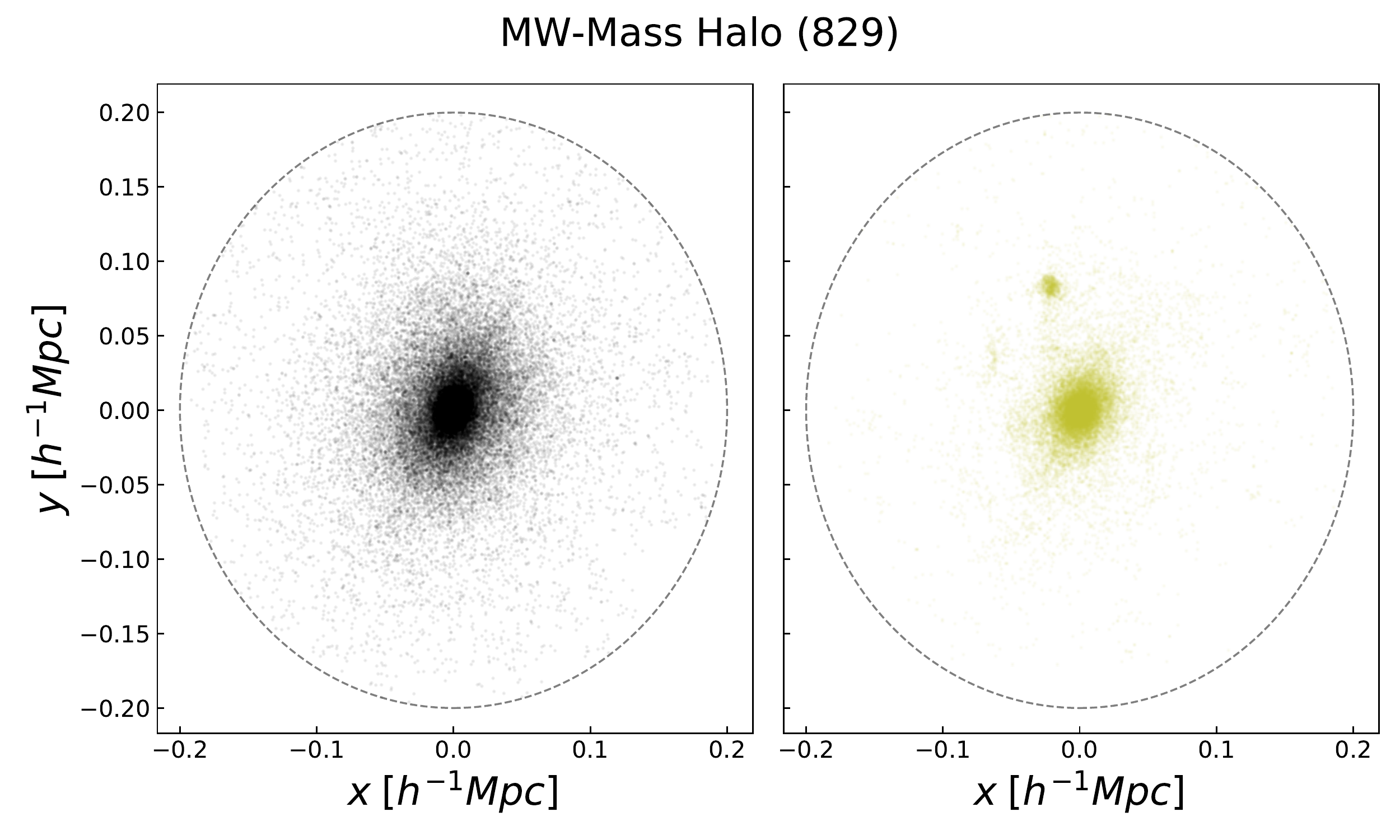}
    \caption{
    \textit{Left}: alternative subhalo-excluded particles, depicted in black. This figure has the same z-axis cut at \autoref{fig:vizualization} and the virial radius of the host halo is overlaid. \textit{Right:} The subset of particles that exist in the fiducial subhalo-excluded method presented in the paper that do not exist in the set of the alternative subhalo-excluded particles. I.e. particles in the middle orange panel of \autoref{fig:vizualization} that do not overlap with the particles shown in the left panel of this plot. It is apparent that in this alternative subhalo-excluded method there is no evidence of other minor over-densities and the host is completely smooth. Additionally, there is a significant portion of mass that does exist in diffuse material that is not yet classified as part of the host halo in our fiducial model.}
    \label{fig:vizualization_appendix}
\end{figure*}

In \autoref{fig:vizualization_appendix} we show a 2D halo slice ($\Delta z = \pm 1$kpc) comparing this alternative subhalo-excluded method to that of \autoref{subsection:calc} and \autoref{fig:vizualization}. The left panel depicts the particles of this alternative subhalo-excluded model in black. The right panel of \autoref{fig:vizualization_appendix} depicts the particles that are in our fiducial model but are not included in the alternative subhaloes excluded model. There is a significant portion of mass in diffuse material that is not yet part of the host halo according to \textsc{Rockstar}, but likely would be in a few time steps. Thus we use this more realistic interpretation as our host halo.

The alternative subhalo-excluded method is an even smoother depiction of the host halo. If we then plotted the opposite of this to show subhaloes only, there would be a much more notable buildup of particles at the centre of the halo caused by the diffuse substructure. This alternative subhalo-excluded method is less physical as haloes would indeed contain diffuse over-density peaks near the centre as a result of hierarchical buildup. Hence we warn the reader about using this alternative subhalo-excluded definition.

\begin{figure*}
    \centering
    \begin{subfigure}{0.5\textwidth}
        \includegraphics[width=0.9\textwidth]{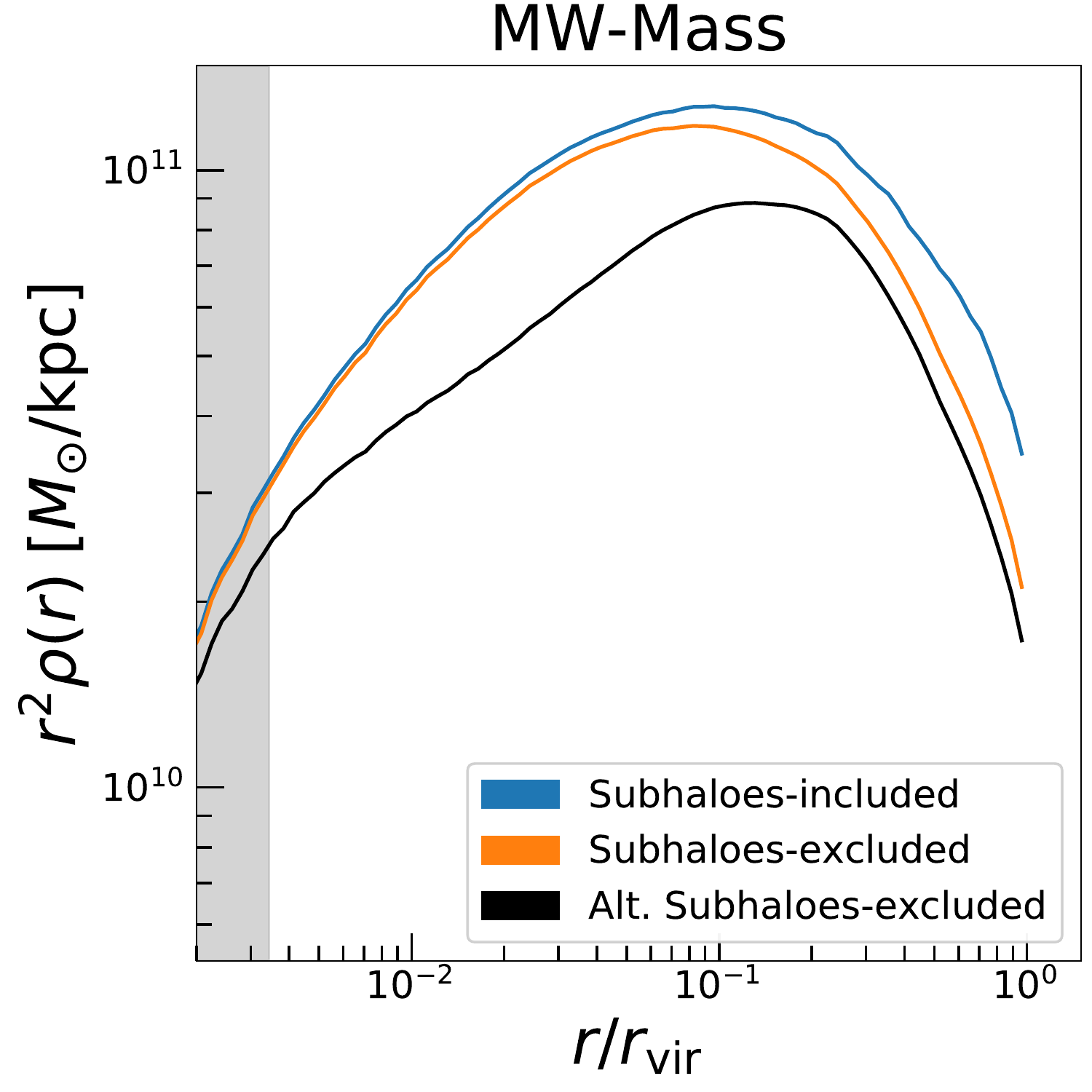}
        \caption{}
        \label{subfig:fullstack_mmmz_appendix}
    \end{subfigure}%
    \begin{subfigure}{0.5\textwidth}
        \includegraphics[width=0.9\textwidth]{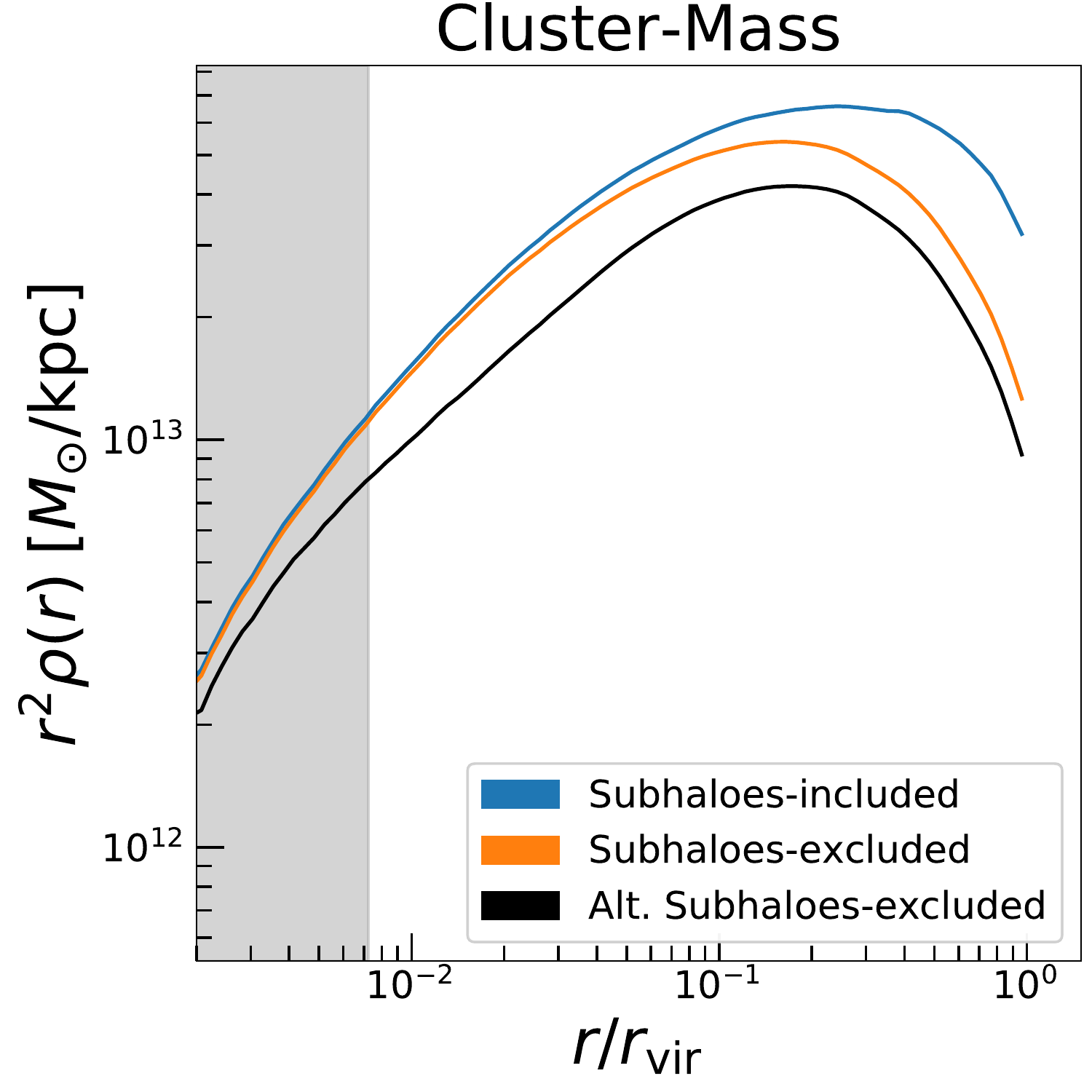}
        \caption{}
        \label{subfig:fullstack_rhapsody_appendix}
    \end{subfigure}%
\caption{\textbf{(a)} The same simulation data depicted in the upper panel of \autoref{subfig:fullstack_mmmz}, i.e. a stacked density profile of all 45 of the MW-mass MMMZ host haloes. The blue curve depicts the density profile for subhalo-included, the orange curve depicts our simulations after subhaloes have been excluded, and the black curve depicts the simulations for the alternative method of full subhalo-excluded. \textbf{(b)} The same data depicted in the upper panel of \autoref{subfig:fullstack_rhapsody}, or a stacked density profile of all 96 of the cluster-mass RHAPSODY host haloes depicting both methods of subhalo-excluded. It is evident that this method excludes a significant portion of mass in the inner halo region in addition to the outer halo, which is likely a less physical representation of an observed halo.} 
\end{figure*}

\autoref{subfig:fullstack_mmmz_appendix} and \autoref{subfig:fullstack_rhapsody_appendix} show what these differences look like when shown as a halo density profile. These plots are identical to the simulation data depicted in our other stacked figures, i.e., the upper panels of \autoref{subfig:fullstack_mmmz} and \autoref{subfig:fullstack_rhapsody}. In addition we provide a stacked profile for the alternative method of subhalo-excluded depicted by the black line. Using this definition of subhalo exclusion more strongly impacts the inner portion of the halo in addition to the outer portion of the halo. This method over-suppresses the inner mass of a realistic halo, as we expect the diffuse/smooth halo component to partially consist of old disrupted subhaloes. 

\section{Best Fit Profile Values}
\label{Section:best_fit_vals}
For reference in \autoref{tab:best_fit} we provide the best fit parameters for the stacked haloes to each profile discussed in the text in addition to the concentration computed from these parameters, as per \autoref{Subsection:methods}. We emphasise that the $r_{s}$ provided is not the inverse of the concentration, as our concentrations are calculated from $r_{-2}$ and not $r_{s}$.

\begin{table}
\centering
\includegraphics[trim=4mm 5mm 5mm 15mm, clip,width=0.47\textwidth]{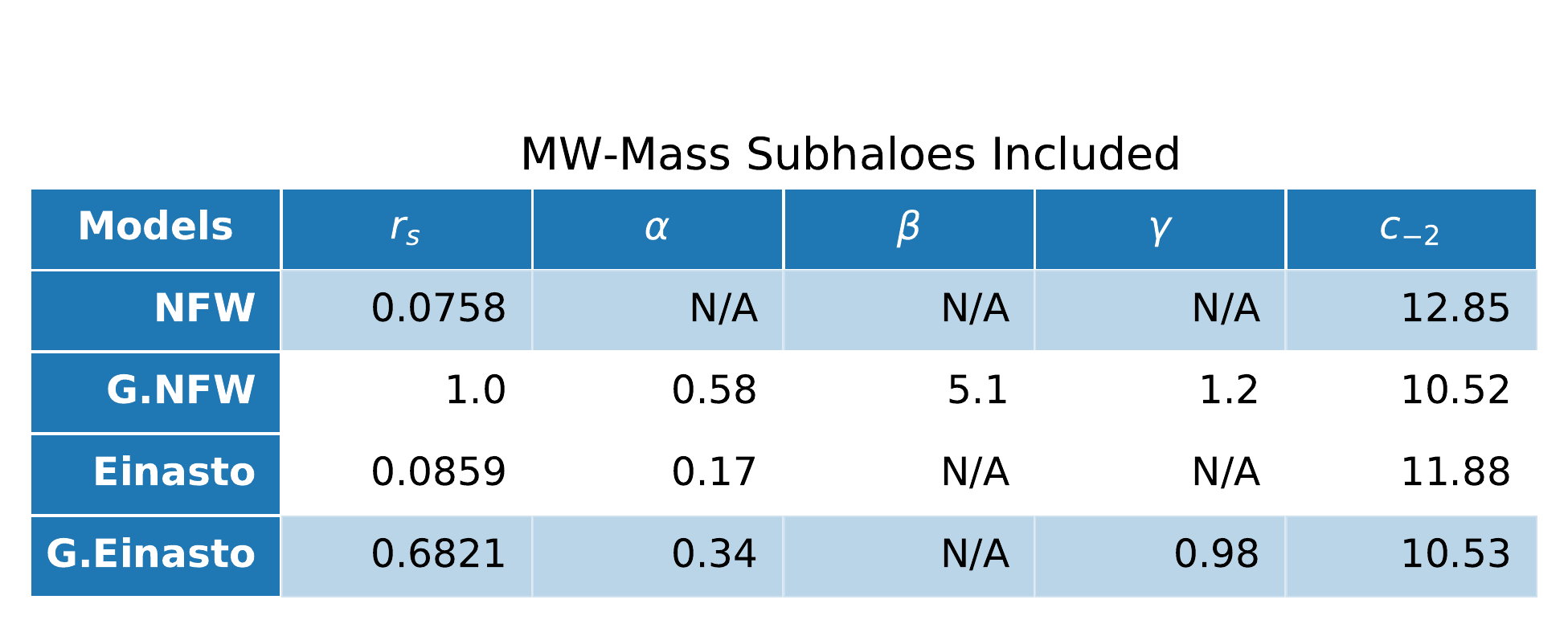}
\includegraphics[trim=4mm 5mm 5mm 15mm, clip,width=0.47\textwidth]{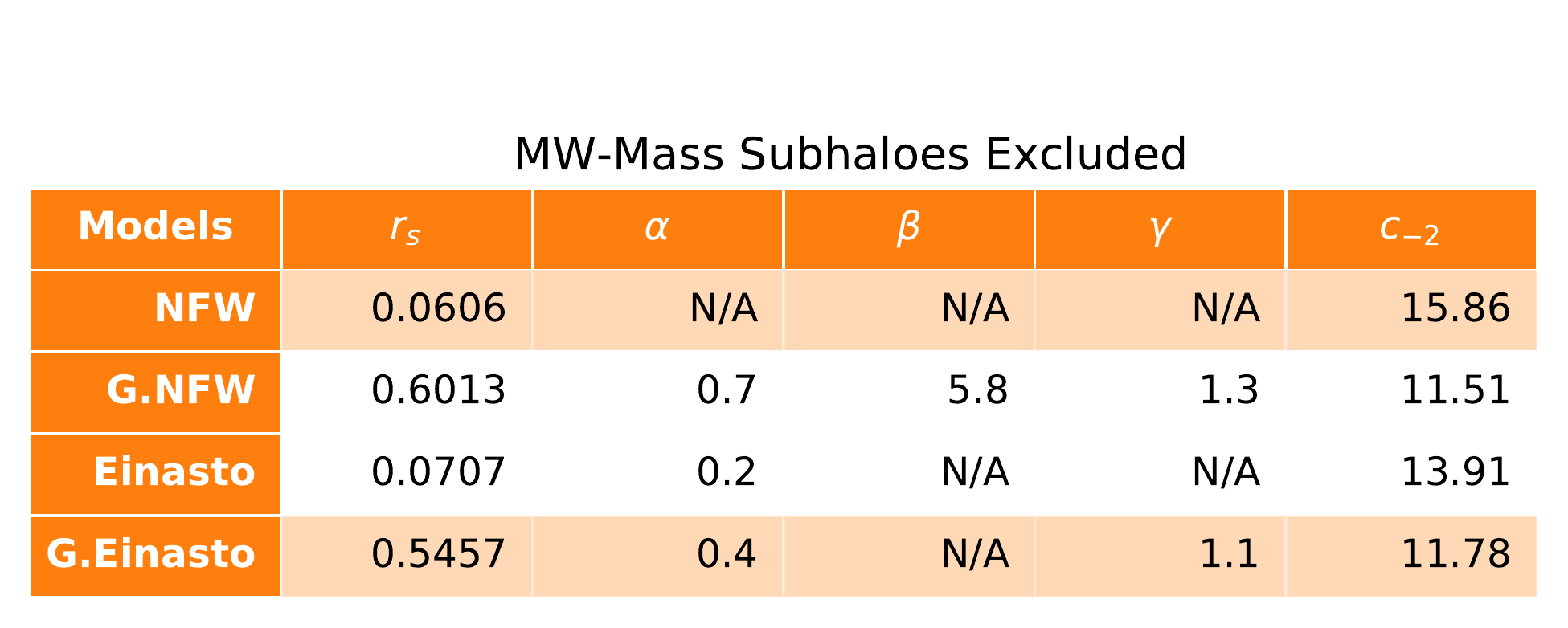}
\includegraphics[trim=4mm 5mm 5mm 15mm, clip,width=0.47\textwidth]{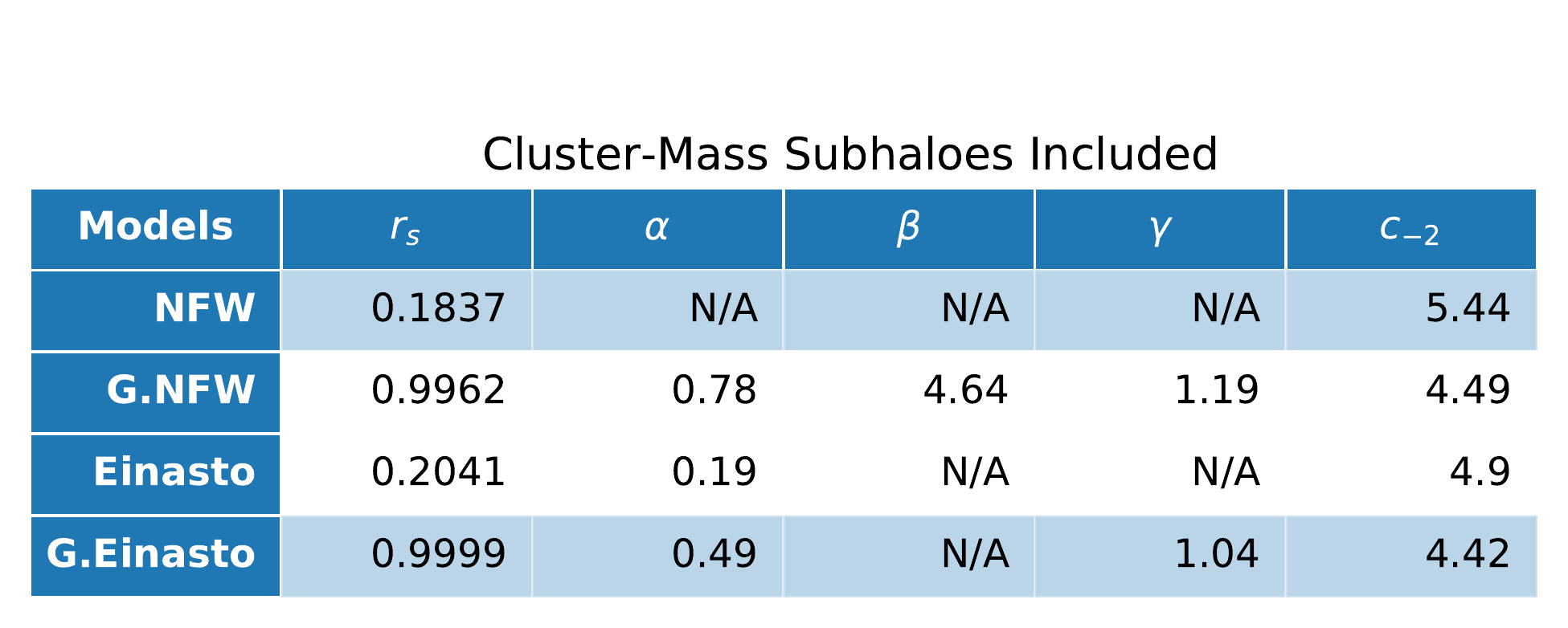}
\includegraphics[trim=4mm 5mm 5mm 15mm, clip,width=0.47\textwidth]{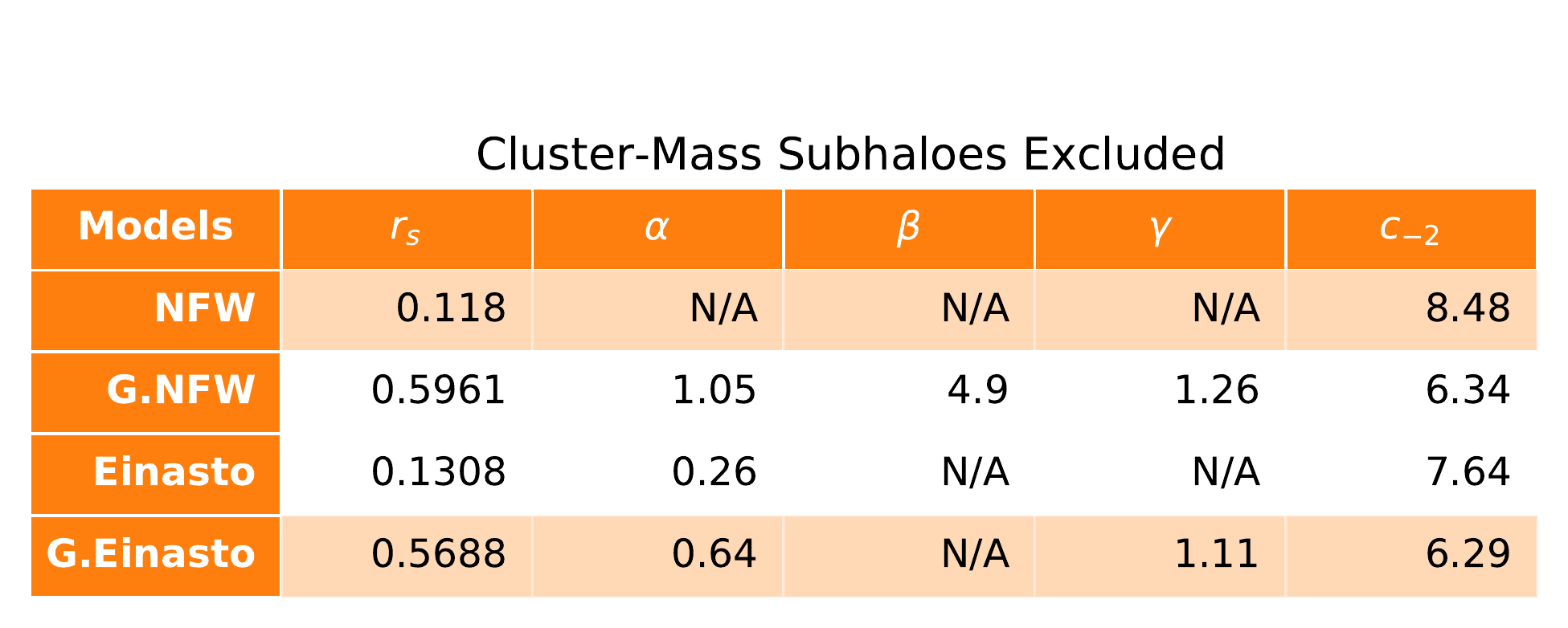}
\caption{Fit values for the profiles of interest in this work, described in \autoref{Subsection:methods}. The $r_{s}$ are scaled by the median $r_{\rm vir}$ as in our plots. These are fit values that result from the stacked haloes after being fit to each listed profile. Values marked as "N/A" mean that this profile does not have that parameter in it.}
\label{tab:best_fit}
\end{table}

\label{lastpage}

\end{document}